\documentclass[fleqn,usenatbib]{mnras}

\usepackage[T1]{fontenc}
\usepackage{ae,aecompl}

\usepackage{graphicx}	
\usepackage{amsmath}	
\usepackage{amssymb}	

\title[The OH Megamaser galaxy IRAS11506-3851]{The OH Megamaser galaxy IRAS11506-3851: an AGN and star formation revealed by multiwavelength observations}

\author[C. Hekatelyne et al.]{
C. Hekatelyne,$^{1}$\thanks{E-mail: hekatelyne.carpes@gmail.com (CH)}
Rogemar A. Riffel,$^{2}$ Thaisa Storchi-Bergmann,$^{1}$ Preeti Kharb,$^{3}$
\newauthor Andrew Robinson,$^{4}$ Dinalva Sales,$^{5}$ Claudia M. Cassanta$^{2}$ 
\\
$^{1}$Departamento de Astronomia, Universidade Federal do Rio Grande do Sul, 91501-970, Porto Alegre, RS, Brazil\\
$^{2}$Departamento de F\'isica, CCNE, Universidade Federal de Santa Maria, 97105-900, Santa Maria, RS, Brazil\\
${^3}$ National Centre for Radio Astrophysics, Tata Institute of Fundamental Research, S. P. Pune University Campus, Post Bag 3,\\ Ganeshkhind, Pune 411 007, India \\
$^{4}$ School of Physics and Astronomy, Rochester Institute of Technology, 84 Lomb Memorial Drive, Rochester, NY 14623, USA \\
$^{5}$ Instituto de Matem\'atica, Estat\'istica e F\'isica, Universidade Federal do Rio Grande, Rio Grande 96203-900, Brazil \\
}

\date{Accepted XXX. Received YYY; in original form ZZZ}

\pubyear{2019}

\begin{document}
\label{firstpage}
\pagerange{\pageref{firstpage}--\pageref{lastpage}}
\maketitle

\begin{abstract}
We present Gemini Multi-Object Spectrograph (GMOS) Integral Field Unit (IFU), Hubble Space Telescope (HST) and Very Large Array (VLA) observations of the OH Megamaser (OHM) galaxy IRAS\,11506-3851. The HST images reveal an isolated spiral galaxy and the combination with the GMOS-IFU flux distributions and VLA data allow us to identify a partial ring of star-forming regions surrounding the nucleus with a radius of $\approx$\,500\,pc. While this ring shows starburst excitation and low velocity dispersion, the region internal to the ring shows higher excitation and velocity dispersion values, with values increasing towards its borders at $\approx$\,240\,pc from the nucleus, resembling a projected bubble. The enhanced excitation and velocity dispersion of this bubble surrounds a 8.5\,GHz radio emission structure, supporting its origin in a faint AGN that is mostly shocking the surrounding gas via a plasma ejection seen in radio at the present stage. This is the fourth of the 5 OHM galaxies we have studied so far (from our sample of 15 OHM) for which GMOS-IFU data indicate the presence of a previously unknown faint AGN at the nucleus, consistent with the hypothesis that OHM galaxies harbor recently triggered AGN.

\end{abstract}




\section{Introduction}

Ultra-Luminous Infrared Galaxies (ULIRGs) are luminous galaxies in the infrared \citep[$L_{IR}\ge10^{12}$$\,$L$_\odot$;][]{soifer87}, most of which are the result of advanced mergers of galaxies \citep[e.g.,][]{sanders96}, and present large amounts of molecular gas in their nuclear region (inner few kpc). Although vigorous star formation is usually seen in these objects, and certainly contributes to the high luminosity of the nucleus, it is not yet clear if and how much can also be contributed by an embedded Active Galactic Nucleus (AGN). Many studies indeed argue that the most luminous AGN (L$\ge 10^{42}\,$L$_\odot$) are preferentially hosted by galaxy mergers \citep[][and references therein]{SB-SM19}.

Many ULIRGs host OH Maser and Megamaser emission, which are the result of stimulated emission appearing at the frequencies 1665 and 1667 MHz lines with luminosities 10$^{2 - 4}$ L$_{\odot}$ \citep{Darling2002,Lo2005}. Galaxies presenting such emission are called OH Megamaser (OHM) galaxies. There is evidence that many of these OHM galaxies host an AGN that is still immersed in dense layers of dust and gas, suggesting that they could represent a key stage in galaxy evolution in which the AGN is being triggered by the accretion of matter to the central supermassive black hole (SMBH). \citep{sanders88,Barnes92,Hopkins2006,Haan2011}. 

Theoretical models and numerical simulations attribute a key contribution from Starbursts and AGN feedback to galaxy evolution \citep{Hopkins2012,Nelson2015,Santaella2018,Schaye2015}. This contribution occurs through the injection of matter and energy into the interstellar medium, originating outflows that play an important role in regulating the growth of the galaxy stellar mass and the accretion of matter to the  central supermassive black hole (SMBH). Such gas outflows are commonly detected in ULIRGS \citep{Arribas2014,Cazzoli2016,Cicone2014}.

This work is part of an ongoing project aimed to investigate the nature of OHM galaxies using Hubble Space Telescope (HST), Very Large Array (VLA) images and Integral Field Spectroscopy (IFS) data. In particular, our goal is to investigate the hypothesis that the megamaser traces a phase in the galaxy in which the AGN is being triggered. We began this study with an initial sample of 15 OHM galaxies selected for being on different merger states -- as it is known that the megamaser is also linked to galaxy mergers and star formation -- in order to investigate if the presence of the AGN is linked to a specific stage of the merger.

In previous papers we discussed the kinematics and excitation of the gas as well as the radio emission in four OHM galaxies. The main results of these works are summarized as follows. \citet{Dinalva2015} performed a multi-wavelength analysis of the OHM galaxy IRAS\,16399-0937, concluding that it harbors an embedded AGN of relatively modest luminosity and that it is also the likely source of the OHM emission observed in this system, while the other nucleus is starburst dominated. In \citet{Heka2018a}, we used Gemini Multi-Object Spectrograph (GMOS) integral field unit (IFU), VLA and HST, which reveal an obscured type 2 AGN and its outflow in the OHM galaxy IRASF\,23199+0123. We also found knots of star formation in the nuclear region of the eastern galaxy of the pair and report a new OH maser detection. In \citet{Heka2018b} we used multi-wavelength data to conclude that IRAS03056+2034 presents evidence of an embedded AGN and circumnuclear star formation that also contributes to the ionization of the gas. Finally, in \citet{Dinalva2019}, multi-wavelength observations show that IRAS\,17526+3253 presents two kinematic components and the emission-line ratio diagnostic diagrams indicate that young stars are the main source of ionization for both of them. Although we find evidence of gas ionized by AGN, the number of objects is still too small to drive a firmer conclusion about the nature of OHM galaxies. 

Here we present multi-wavelength observations of another OHM galaxy, IRAS\,11506-3851 (hereafter IRAS\,11506, also identified as ESO 320-G030), which is an isolated spiral galaxy \citep{Bellocchi2013,Bellocchi2016} showing a double-barred structure \citep{Greusard2000} with no clear evidence of having suffered a major merger and hosting a weak maser (\citet{Norris1986}), what is consistent with results from previous studies that show that major mergers host more powerful OHMs.

The weak maser in IRAS11506 shows two components: a stronger OH feature accompanied by a weaker, and slightly blueshifted one. Each feature has a half-power width of $\approx$ 120 km $s^{-1}$, which is typical of OHM. Moreover, considering H$_{0}$=75 km s$^{-1}$ Mpc$^{-1}$, they estimated that the isotropic luminosity of the OH emission is $\approx$ 52 L$_{\odot}$\,\citep{Norris1986}.

IRAS\,11506 is a local (46 Mpc; scale of 240 pc arcsec$^{-1}$, \citet{Cazzoli2014}) LIRG (log L$_{IR}/L_{\odot}$=11.3) classified as presenting an HII (or Starburst) nucleus from optical spectroscopy \citep{Van1991}. Previous studies reveal a resolved outflow in neutral \citep{Cazzoli2014} and molecular gas using Atacama large Millimetric Array (ALMA) observations \citep{Santaella2016}.

We use spectroscopic data obtained with the GMOS--IFU, combined with VLA and HST images of IRAS11506 to map the stellar and gas kinematics and distributions, as well as the gas excitation in the inner kpc of the galaxy. This paper is organised as follows: in Section 2 we describe the observations and data reduction. Section 3 presents our results, in Section 4 we discuss the nature of the nuclear source and gas kinematics and we present our conclusions in Section 5.

\section{Observations and data reduction}

\subsection{HST data}

\begin{figure*}
	\includegraphics[width=\textwidth]{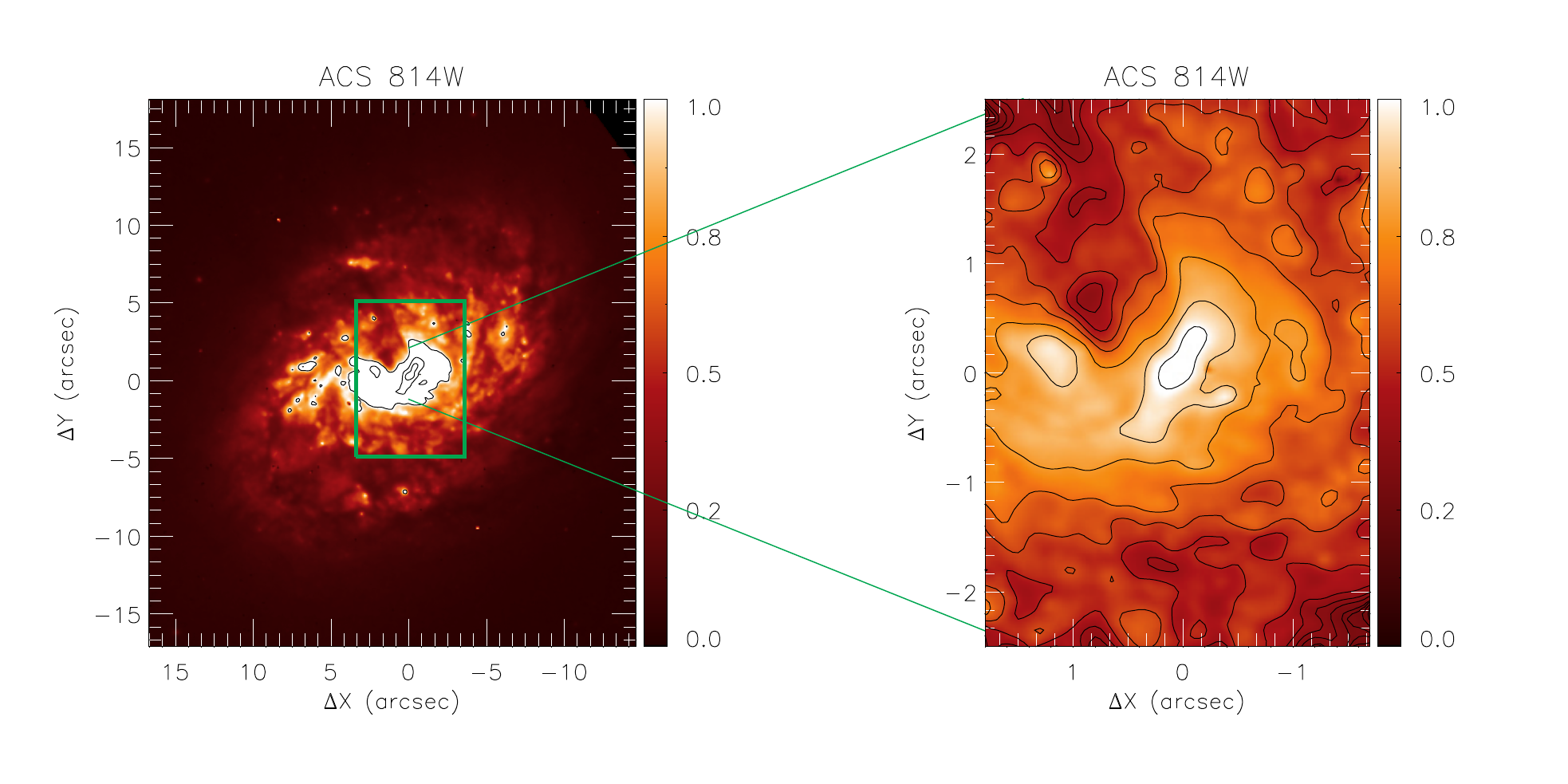}
	\includegraphics[width=\textwidth]{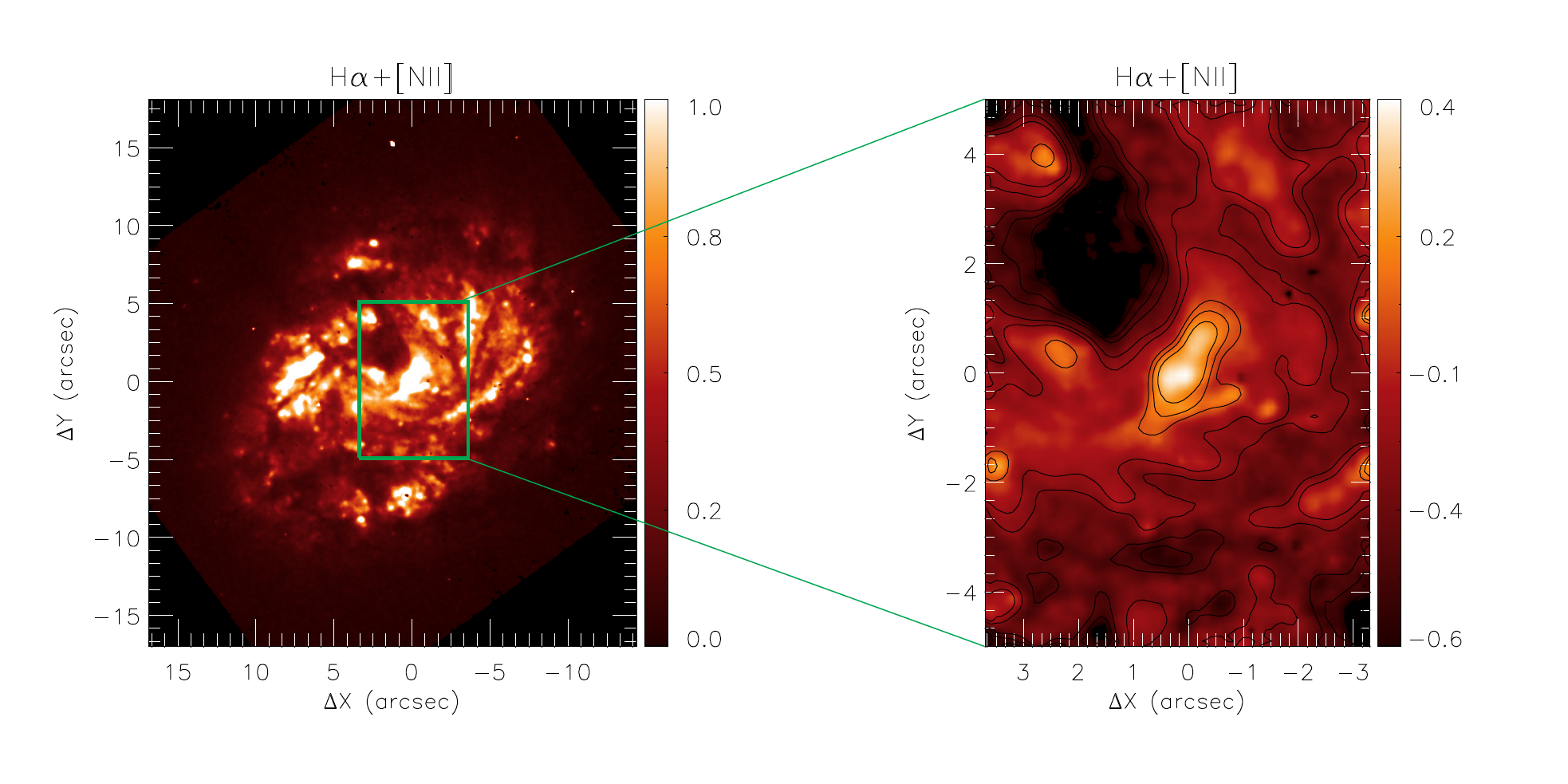}
    \vspace{-0.5cm}
    \caption{HST images of IRAS\,11506. Left panels: Large-scale images, showing in the top panels the ACS/HST F814W - i band images and, in the bottom panels, the H$\alpha+$[N\,{\sc ii}]$\lambda\lambda6548,84$ narrow-band images.  Right panels: zoom-in of the images in the left showing the region observed with GMOS-IFU, whose FoV (3\farcs5$\times$5\farcs5) is shown as the green rectangles. The color bars show the fluxes in arbitrary units.}
    \label{fig:hst}
\end{figure*}

HST continuum and emission-line imaging of 15 OHM galaxies were obtained in Program 1160 (PI  D. J. Axon), including IRAS\,11506. The data were obtained with the Advanced Camera for Surveys (ACS) using wide, broad, narrow and medium-band filters. The reduction process was performed using Image Reduction and Analysis Facility ({\sc iraf}) \citep{tody86,tody93} routines and followed the basic procedures described in \citet{Dinalva2019}. Figure\,\ref{fig:hst} shows the ACS F814W continuum and narrow-band images in H $\alpha+$[N\,{\sc ii}]$\lambda\lambda6548,84$ emission lines, highlighting the observed structure within the GMOS-IFU field-of-view (hereafter FoV).

\subsection{VLA radio continuum data}

We reduce 1.43 and 4.86 GHz archival data acquired with the VLA in the A-array configuration and 8.46 GHz data in the AB-array configuration (AL508) using standard calibration and imaging procedures in {\sc aips}. A couple of rounds of phase-only and phase+amplitude self-calibration were carried out interactively using the {\sc aips} task {\sc calib} before producing the final radio images. The {\it rms} noise in the 1.43 and 4.86 GHz images are 0.1 mJy/beam and 90 mJy/beam, respectively. We create an 1.43 - 4.86 GHz spectral index image using the task {\sc comb} after convolving the images at the two frequencies with a circular beam of 2\farcs5; flux density values below the 3 sigma level were blanked. Figures\,\ref{fig:radio} and \ref{fig:spectralindex} show the radio continuum image at 8.5 GHz and the 1.43-4.86 GHz spectral index image, respectively.

\subsection{GMOS-IFU data}

IRAS11506 was observed at the Gemini South Telescope GMOS-IFU  \citep{allington-smith02}. The observations were carried out on the nights of 2014 April 20, 21 and 2014 May 08,18 splitted into 23 individual exposures, 17 of them having exposure times of 1200 sec and 6 of them of 1100 sec. 

The observations were done with the IFU operating in the one-slit mode using the IFU-R mask, with the grating B600$_-$G5323 and the filter GG455$_-$G0329. We obtain spectra centred at three distinct wavelengths (6200\AA, 6300\AA and 6400\AA) to properly correct for the gaps between the detectors and we orient the IFU along position angle 358$^\circ$, chosen to find suitable guide stars. This configuration corresponds to a FoV of 3\farcs5$\times$5\farcs0, and spectral range is such that it includes the strongest optical lines, from H$\beta$ to [S\,{\sc ii}]$\lambda$6731\AA.

The processing of the data was performed using the {\sc gemini iraf} package following the standard procedures of spectroscopic data reduction. The main steps include the subtraction of the bias level for each image, flat-fielding and trimming. We then perform the wavelength calibration using the spectra of CuAr lamps as reference, followed by the subtraction of the sky emission. The sensitivity function was obtained from the spectrum of the EG\,131 standard star in order to flux calibrate the spectra. We finally obtain a data cube for each exposure, which were then median combined using the {\sc gemcombine} routine, resulting in a single data cube for the object. During this step we used the peak of the continuum emission as reference and the {\sc sigclip} algorithm to remove bad pixels and remaining cosmic-ray contamination.

The seeing during the observation is $\sim$0\farcs6, as obtained from the measurement of the full width at the half maximum (FWHM) of the flux distribution of the standard star EG131. This translates to $\sim$115\,pc at the galaxy considering a distance of 46\,Mpc \citep{Cazzoli2014}. The velocity resolution is $\sim$80\,kms$^{-1}$, as estimated from the measurement of the FWHM of typical emission lines of the CuAr lamps spectra.

\section{RESULTS}

\subsection{HST images}

In Figure \ref{fig:hst}, we present in the ACS/HST F814W i-band image (top panel) and the narrow-band H${\alpha}$+[N\,{\sc ii}]$\lambda\lambda6548,84$ image (bottom panel) of IRAS11506. The panels on the left-hand show large scale images of the galaxy and the green boxes represent the FoV of the GMOS-IFU data. The right-hand panels present zoomed images of the galaxy within the same FoV of the GMOS-IFU data.

The i-band image reveals the spiral structure of the galaxy with the highest fluxes observed at its nucleus. The zoomed i-band image shows a region of fainter emission to the north-east of the nucleus, possibly due to a higher dust extinction in that region. In addition, several knots of emission are seen surrounding the nucleus, with the brightest structure resembling a partial ring or segments of spiral arms.

The continuum-free HST H${\alpha}$+[N\,{\sc ii}]$\lambda\lambda6548,84$ narrow-band image presents a flux distribution similar to that of the i-band image, with the strongest emission along PA$\approx135/345^\circ$ and several knots of emission associated to regions with the enhanced i-band continuum fluxes.

\subsection{VLA images}

\begin{figure}
\vspace{-1cm}
\includegraphics[width=0.45\textwidth]{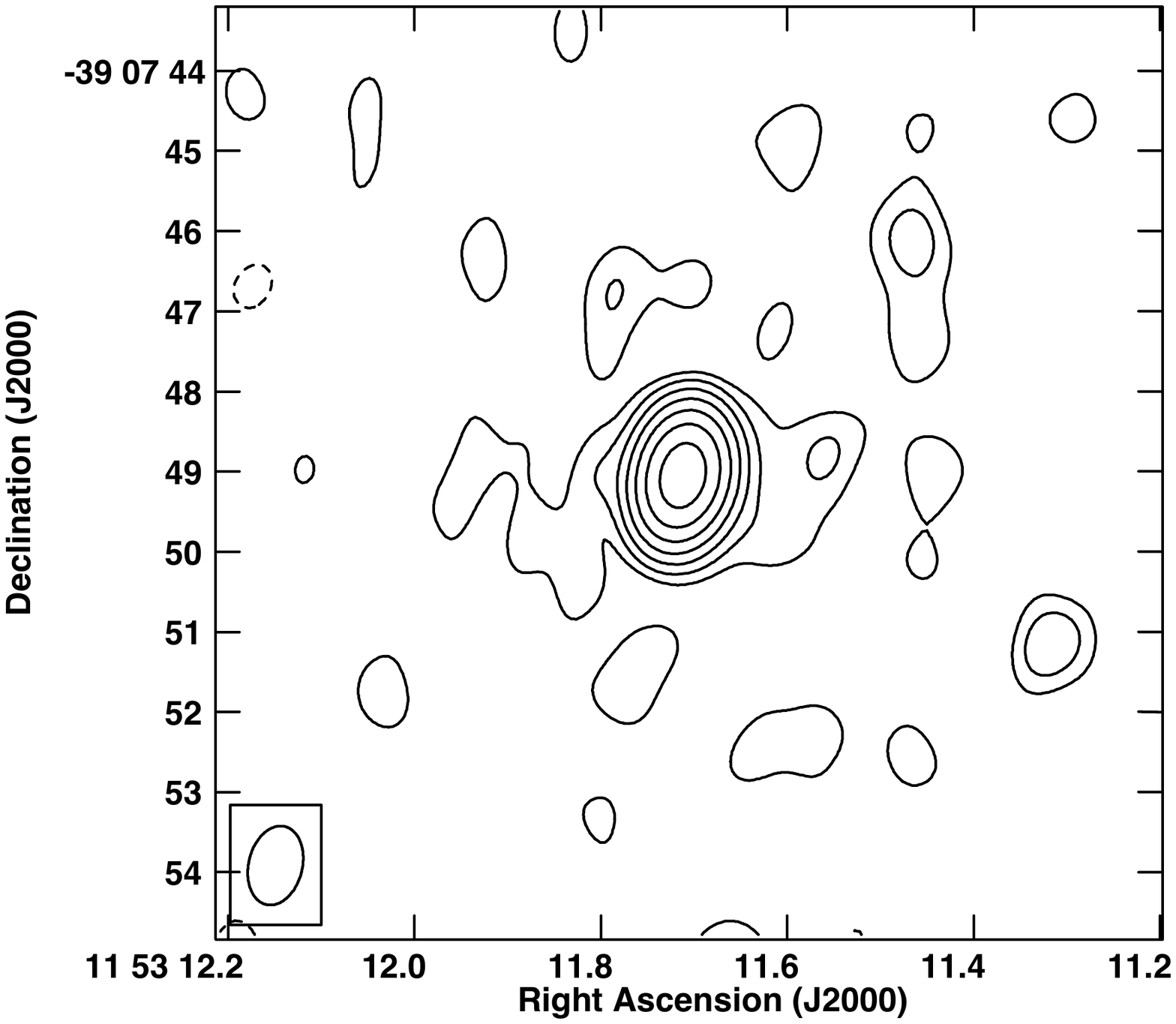}
\vspace{-2.0cm}
\caption{VLA radio continuum images at 8.5 GHz.
The peak surface brightness and lowest contour level is 25.9 mJy/beam and $\pm$ 1\%, respectively. The beam is $1.00\times0.67$ arcsec at a PA=-13.5 degrees.}
\label{fig:radio}
\end{figure}

Figure~\ref{fig:radio} presents the VLA continuum image at 8.5\,GHz that reveals a two-sided core-jet-like structure in the east-west direction of extent ~5\farcs5. This extension is better defined in the 1.4-4.9\,GHz spectral index image which shows a steep spectral shape (spectral index of -0.80$\pm$0.11) along the same direction. The core spectral index is relatively flatter at -0.54$\pm$0.01.

\begin{figure}
	\hspace{-1.2cm}
	\includegraphics[width=0.60\textwidth]{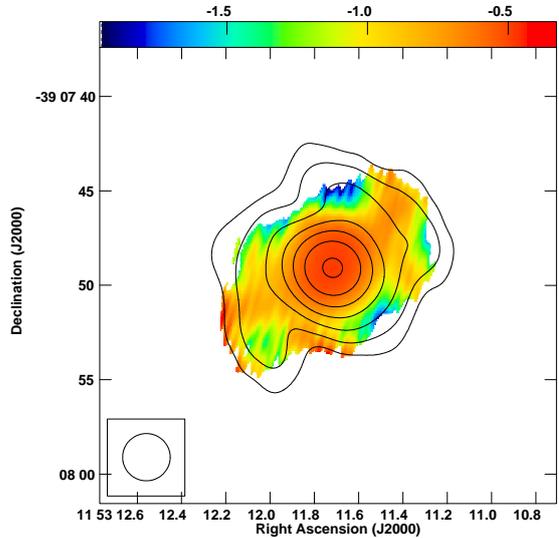}
    \vspace{-0.5cm}
    \caption{Radio spectral index image of IRAS11506, derived from 1.4 and 4.9 GHz VLA images at a resolution of 2.5 arcsec. The contours are at 1.4 GHz with the lowest contour level and peak surface brightness value being +/- 1.4\% and 5.8E-2 Jy/beam.}
    \label{fig:spectralindex}
\end{figure}

An elongation is clearly seen in the 1.4-4.9 GHz spectral index map of Figure~\ref{fig:spectralindex} actually more along the southeast-northwest direction (instead of east-west direction), while the highest spectral index values of $-0.5$ are found at the nucleus.

\begin{figure*}
\includegraphics[width=\textwidth]{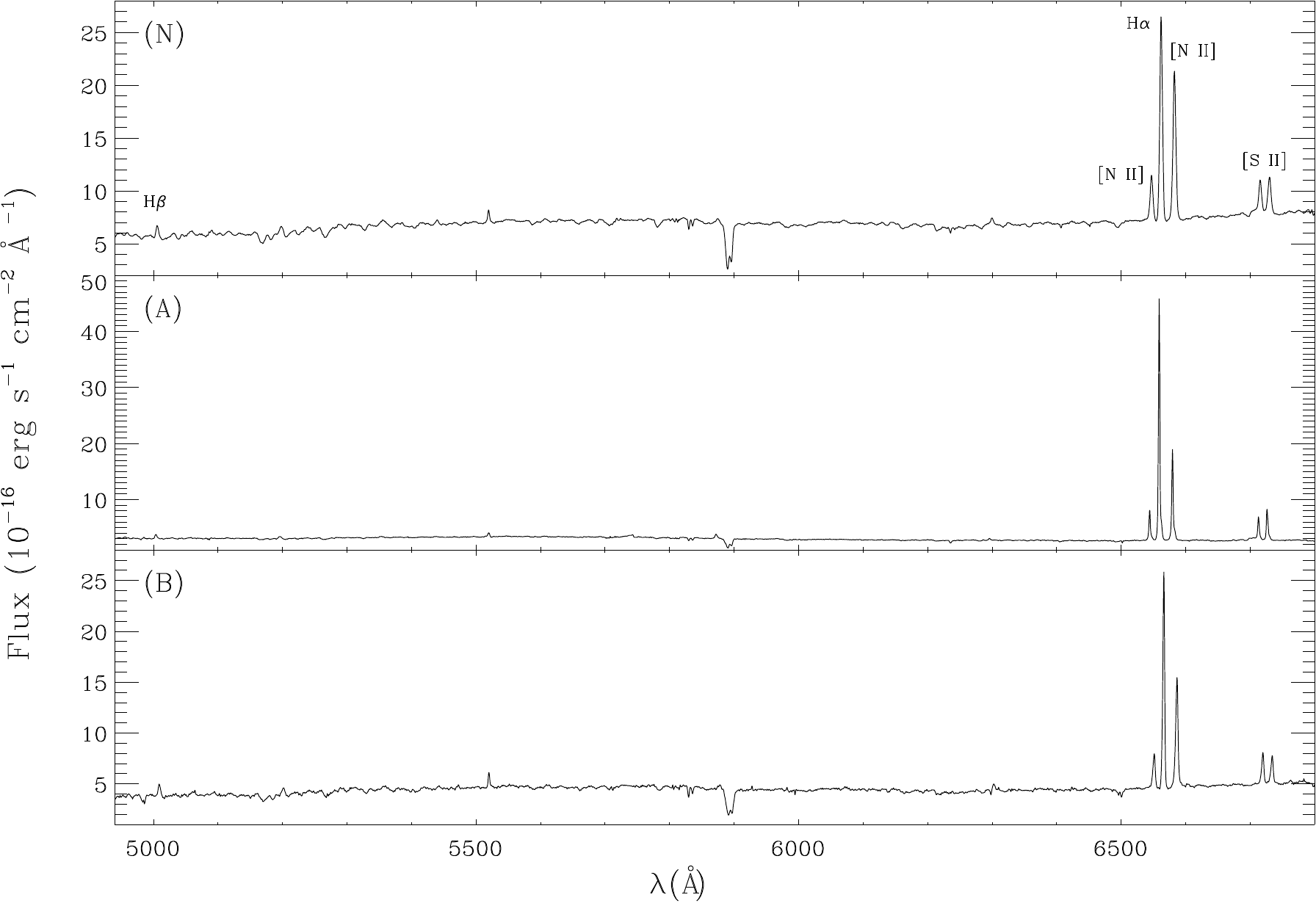}
\caption{Three representative spectra of the different regions probed by our observations, from top to bottom: the nucleus (N in Fig.\ref{fig:flux}), a region between the nucleus and the ring (A in Fig.\ref{fig:flux}) and a region in the ring (E in Fig.\ref{fig:flux})}
\label{fig:spectra}
\end{figure*}

\subsection{GMOS-IFU Emission-line flux distributions}

\begin{figure*}
	\includegraphics[width=\textwidth]{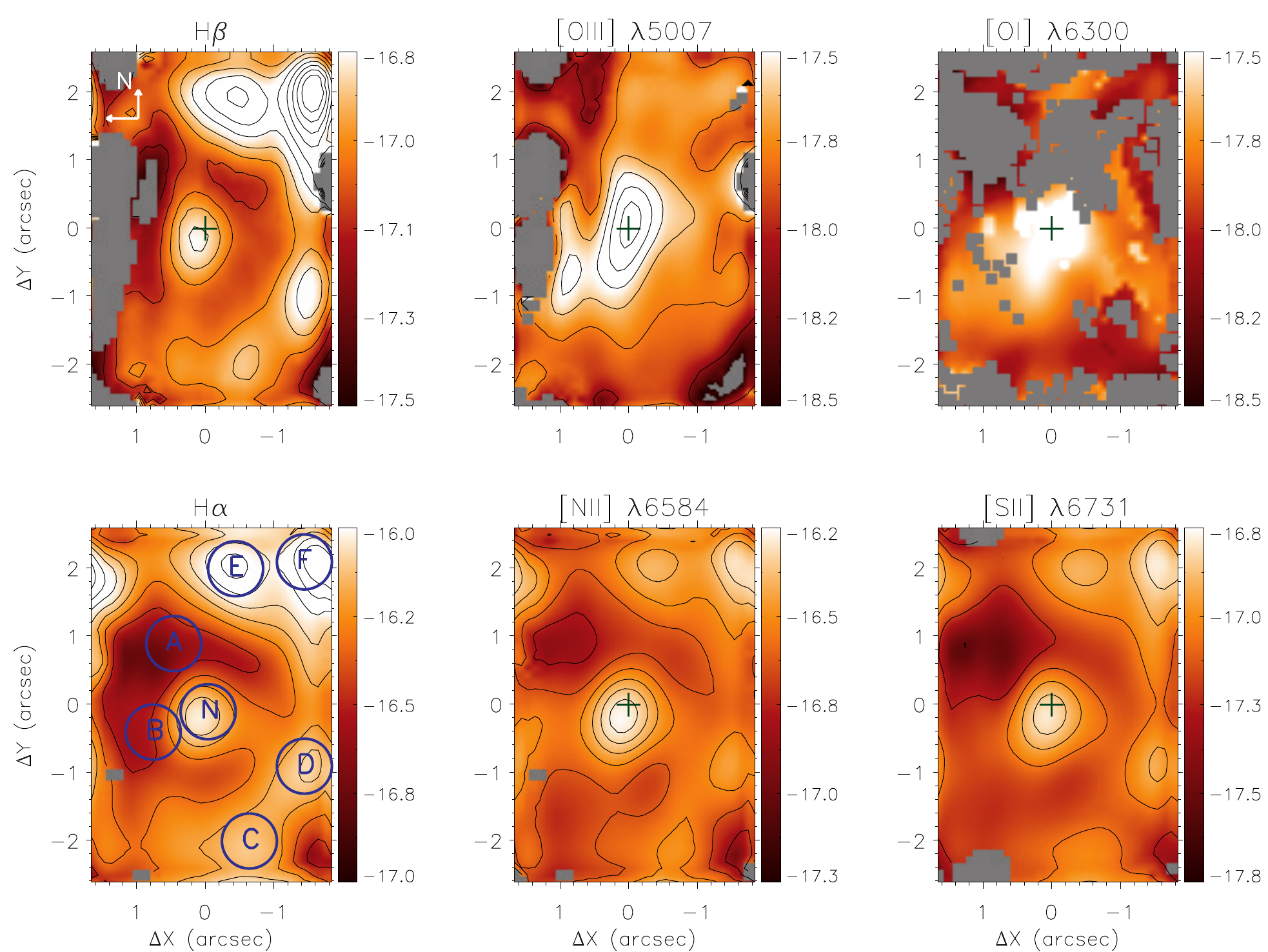}
    \caption{Top panels: flux maps for H$\beta$ (left), [O\,{\sc iii}]$\lambda$5007 (centre) and [O\,{\sc i}]$\lambda$6300 (right) emission-lines of IRAS11506. Bottom panels: flux maps for H$\alpha$ (left), [N\,{\sc ii}]$\lambda$6584 (centre) and [S\,{\sc ii}]$\lambda$6731 (right). The central crosses in all maps mark the position of the nucleus and grey regions represent masked locations, where the signal-to-noise ratio was not high enough to obtain reliable fits of the emission-line profiles or locations with no line detection. The blue circles in the H$\alpha$ flux map delimit the regions where we have extracted spectrum to characterize the star-forming properties (see Fig.\, \ref{fig:spectra}). The color bars show the fluxes in logarithmic units of erg\,s$^{-1}$cm$^{-s}$spaxel$^{-1}$. }
    \label{fig:flux}
\end{figure*}

We have mapped the emission-line flux distributions of H$\alpha$, H$\beta$, [O\,{\sc iii}]$\lambda$5007, [O\,{\sc i}]$\lambda$6300, [N\,{\sc ii}]$\lambda$6584 and [S\,{\sc ii}]$\lambda$ 6731 (hereafter [O\,{\sc iii}], [O\,{\sc i}], [N\,{\sc ii}] and [S\,{\sc ii}], respectively) lines by fitting the line profiles by Gaussian curves. The fitting is done using a modified version of the {\sc profit} routine \citep{profit}, adopting a similar fitting procedure as described in \citet{Izabel}. In short, each line is fitted by a single Gaussian component and the underlying continuum is represented by a linear equation fitted to continuum regions, next to each line profile. We fit the H$\beta$+[O\,{\sc iii}] line complex simultaneously, tying the width and velocity of the  [O\,{\sc iii}]$\lambda$4959 and [O\,{\sc iii}]$\lambda$5007 emission lines and 
keep fixed their flux ratio to the theoretical value of  [O\,{\sc iii}]$\lambda$5007/[O\,{\sc iii}]$\lambda$4959$=3$ \citep{osterbrock}. A similar procedure is adopted for the H$\alpha$+[N\,{\sc ii}] complex, where we tie the kinematics of the [N\,{\sc ii}] lines and  [N\,{\sc ii}]$\lambda6583$/[N\,{\sc ii}]$\lambda6548=3$. The kinematics of the [S\,{\sc ii}] lines is also tied, while the [O\,{\sc i}] is fitted separately.

Figure \ref{fig:spectra} shows three representative spectra of different regions probed by our observations: the nucleus (N in Fig.\ref{fig:flux}), a region between the nucleus and the ring of star forming regions (A in Fig.\ref{fig:flux}) and a region in the ring (E in Fig.\ref{fig:flux}).

Figure~\ref{fig:flux} shows the H$\beta$, [O\,{\sc iii}] and [O\,{\sc i}] flux maps  in the top panels and for the H$\alpha$, [N\,{\sc ii}] and [S\,{\sc ii}] in the bottom panels. The central cross in all maps indicates the position of the nucleus (identified with the peak of continuum emission) and grey regions represent masked locations where the signal-to-noise ratio was not high enough to allow reliable measurements of the emission lines or locations with no detection of line emission. These locations present flux uncertainty larger than 30 per cent. The color bars show the fluxes in logarithmic units of erg s$^{-1}$cm$^{-s}$spaxel$^{-1}$.

The H$\alpha$ and H$\beta$ emission-line maps reveal the strongest emission at $\sim$2\farcs0 north-northwest from the nucleus, in regions identified as E and F in the H$\alpha$ map of Fig.\,\ref{fig:flux}. These regions are part of a partial ring of knots of enhanced emission seen at $\approx$\,2$^{\prime\prime}$ ($\approx$\,500\,pc) from the nucleus, which may be due to circumnuclear  star-forming regions (CNSFRs). For simplicity, we will call this structure hereafter, as the "ring''. The nucleus shows a somewhat lower emission, as compared to the CNSFRs and the lowest emission is seen to the northeast of the nucleus. 

In the [N\,{\sc ii}] and [S\,{\sc ii}] flux maps the nucleus is somewhat more luminous than the brightest CNSFRs of the ring (regions E and F), while in the [O\,{\sc iii}] and [O\,{\sc i}] maps the nucleus is much more luminous than the ring, indicating a distinct nature for the gas excitation of the nucleus and the ring.  In the case of the [O\,{\sc iii}] flux map, the nuclear emission is the most extended among all lines, and delineates an elongated structure of $\approx$1.5$^{\prime\prime}$ (360\,pc) running from the south-east to the north-west. Another region of enhanced [O\,{\sc iii}] emission is observed at $\approx$1\farcs3 (270\,pc) east to south-east of the nucleus, which is not seen in the other emission-line maps.

\subsection{Emission-line ratio maps}
\label{sec:lineratios}

\begin{figure*}
	\includegraphics[width=\textwidth]{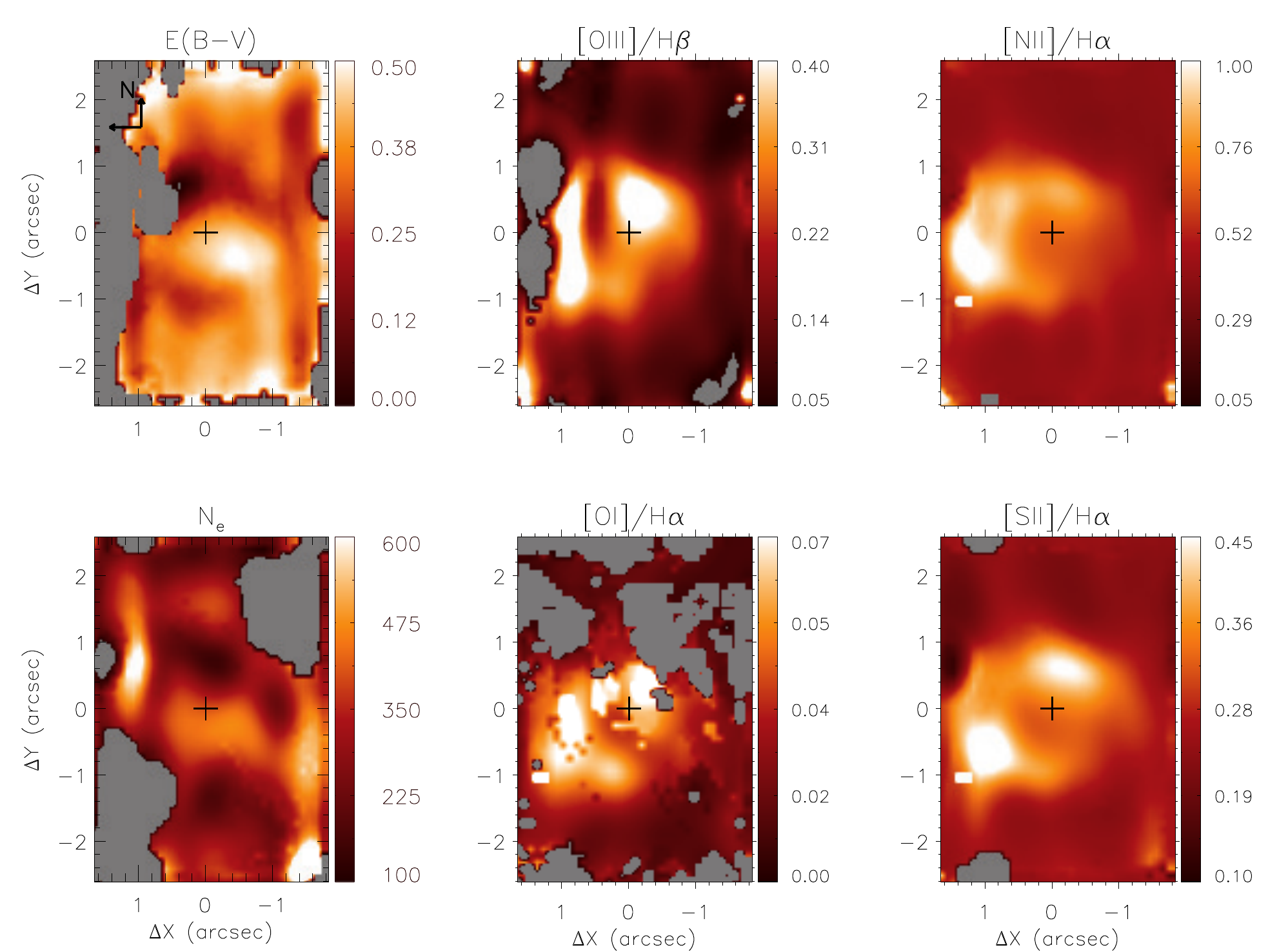}
    \caption{Top panels: E(B-V) map obtained from the H$\alpha$/H$\beta$ line ratio (left), emission-line ratios of [O\,{\sc iii}]/H$\beta$ (centre) and [N\,{\sc ii}]/H$\alpha$ (right). Bottom panels: Electron density map for IRAS11506 obtained from the [S\,{\sc ii}] emission-lines, which the colour bar shows the N$_{e}$ values in units of cm$^{-3}$ (left), emission-line ratios of [O\,{\sc i}]/H$\alpha$ (centre) and [S\,{\sc ii}]/H$\alpha$ (right). }
    \label{fig:ratio}
\end{figure*}

From the emission-line flux distributions, we obtained the reddening and electron density $N_e$ distributions, shown in Figure \ref{fig:ratio} together with the line ratio maps [O\,{\sc iii}]/H$\beta$, [N\,{\sc ii}]/H$\alpha$, [O\,{\sc i}]/H$\alpha$ and [S\,{\sc ii}]$\lambda\lambda6717,6731$/H$\alpha$.

The color excess map [$E(B-V)$] is obtained from the H$\alpha$/H$\beta$ flux ratio following the procedure described in \citet{Izabel}. The values of $E(B-V)$ range from 0.1 to 0.5, with the highest ones observed to the south-west of the nucleus. 

The ionised gas density map $N_e$ was obtained from the [S\,{\sc ii}]$\lambda$6717/6731 line ratio assuming an electronic temperature $T_{E}$=10.000~K using the \textit{temden} {\sc iraf} routine. The map shows values ranging from 100 to 500 cm$^{-3}$ with the highest ones observed to the south and south-west of the nucleus and at the borders of the FoV. 

The [O\,{\sc iii}]/H$\beta$ ratio map shows the lowest values ($\le$0.1) at the ring and the highest values (of up to 0.6) in a knot extending to 1\arcsec (240\,pc) north-west of the nucleus and in an elongated region at 1$\farcs$3 (312\,pc) east and south-east of it, which together with the surrounding values inside the ring show a structure that seems to delineate the borders of a $\approx$\,500\,pc diameter bubble somewhat off-centered relative to the nucleus.

The [N\,{\sc ii}]/H$\alpha$ and [S\,{\sc ii}]/H$\alpha$ maps show the highest values at locations where the [O\,{\sc iii}]/H$\beta$ and [N\,{\sc ii}]/H$\alpha$ are also higher than in the surroundings, better defining the off-centered bubble described above. Although more noisy, the [O\,{\sc i}]/H$\alpha$ ratio map also shows this bubble.

\subsection{Stellar kinematics}

We use the penalized Pixel-Fitting pPXF routine \citep{Cappellari2004,cappellari17} for fitting the absorption spectra of IRAS11506 and measure the stellar kinematics. To perform the measurements we use as spectral templates, the Single Stellar Populations (SSP) models of \citet{bc03}, which have similar spectral resolution of the GMOS data and have been previously used to measure the stellar kinematics of nearby galaxies from GMOS-IFU data \citep{brum17}. The line-of-sight velocity distribution of the stars is assumed to be Gaussian, as we fit only the two first moments ($V_\star$ and $\sigma_\star$). We fit the spectral range from 5050 to 6100\,\AA, masking out emission lines and regions where no absorption lines are present, allowing the use of multiplicative polynomials of order 4 to correct the shape of the continuum and we use the {\it clean} parameter to reject all spectral pixels deviating more than 3$\sigma$ from the best fit, to exclude possible remaining spurious features.

\begin{figure}
	\includegraphics[width=0.45\textwidth]{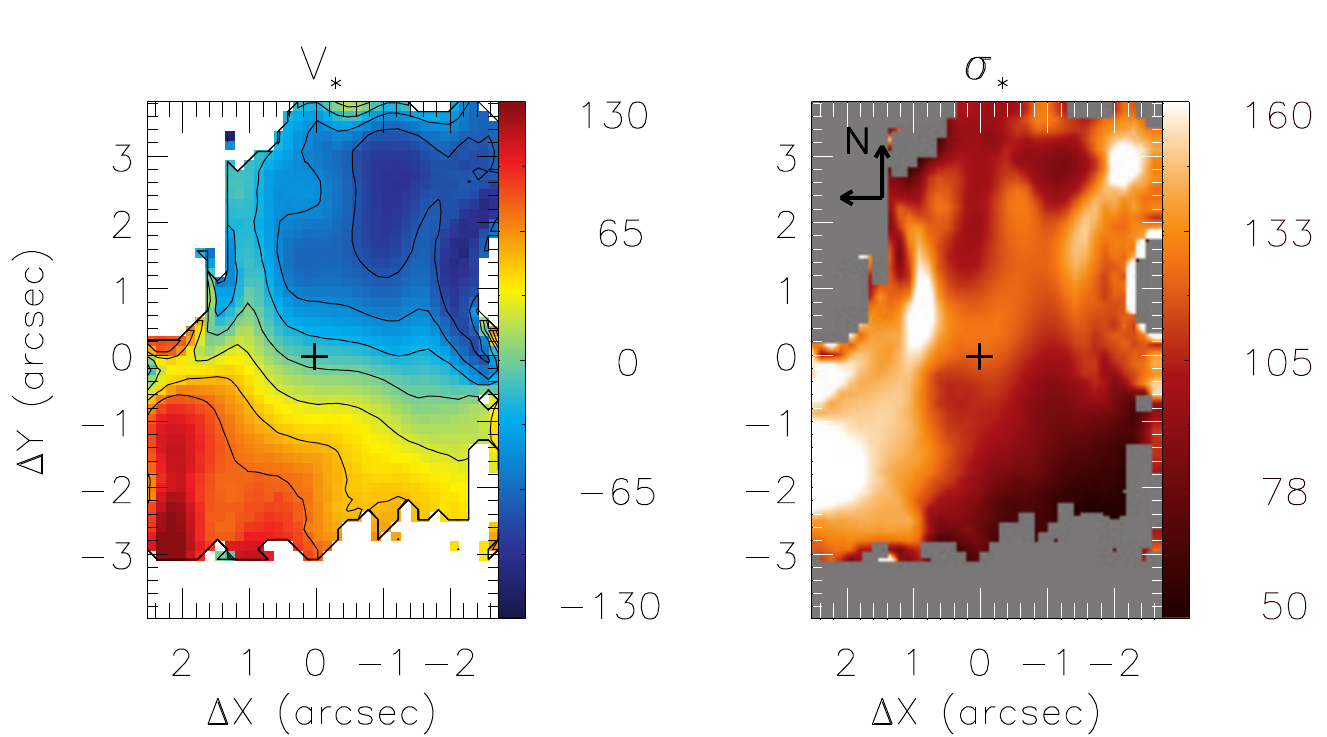}
    \caption{Stellar centroid velocity field and stellar velocity dispersion for IRAS11506. The color bars are in units of km\,s$^{-1}$. The central crosses mark the location of the nucleus.}
    \label{fig:stellar}
\end{figure}

The pPxF routine provides as output the measurements of the stellar velocity V$_{\star}$ and velocity dispersion $\sigma_\star$ as well as their corresponding uncertainties. Figure~\ref{fig:stellar} shows the corresponding two-dimensional maps. Locations in white in the velocity field and in gray in the velocity dispersion map are masked locations, corresponding to spaxels where the uncertainties in V$_{\star}$ and/or $\sigma_\star$  are larger than 30\,km\,s$^{-1}$.  
 The systemic velocity  of the galaxy (3076\,km\,s$^{-1}$), corrected for the heliocentric rest frame is subtracted from the observed velocity, as obtained from the fitting of the rotation disk model discussed in Sec.\,\ref{sec:kinematics}. The stellar velocity field shows a well defined rotation pattern with blueshifts to the north-west and redshifts to the south-east with a projected velocity amplitude of about 120 km\,s$^{-1}$. 

The stellar velocity dispersion map presents the larger values (of $\approx$\,140\,km\,s$^{-1}$) along $PA\approx135/315^\circ$, coincident with the orientation of the elongated structure seen in the HST continuum image (Fig.~\ref{fig:hst}). The smallest values of 70--100 km\,s$^{-1}$ are observed mainly to the southwest of the nucleus.

\subsection{Gas Kinematics}

\begin{figure*}
	\includegraphics[width=\textwidth]{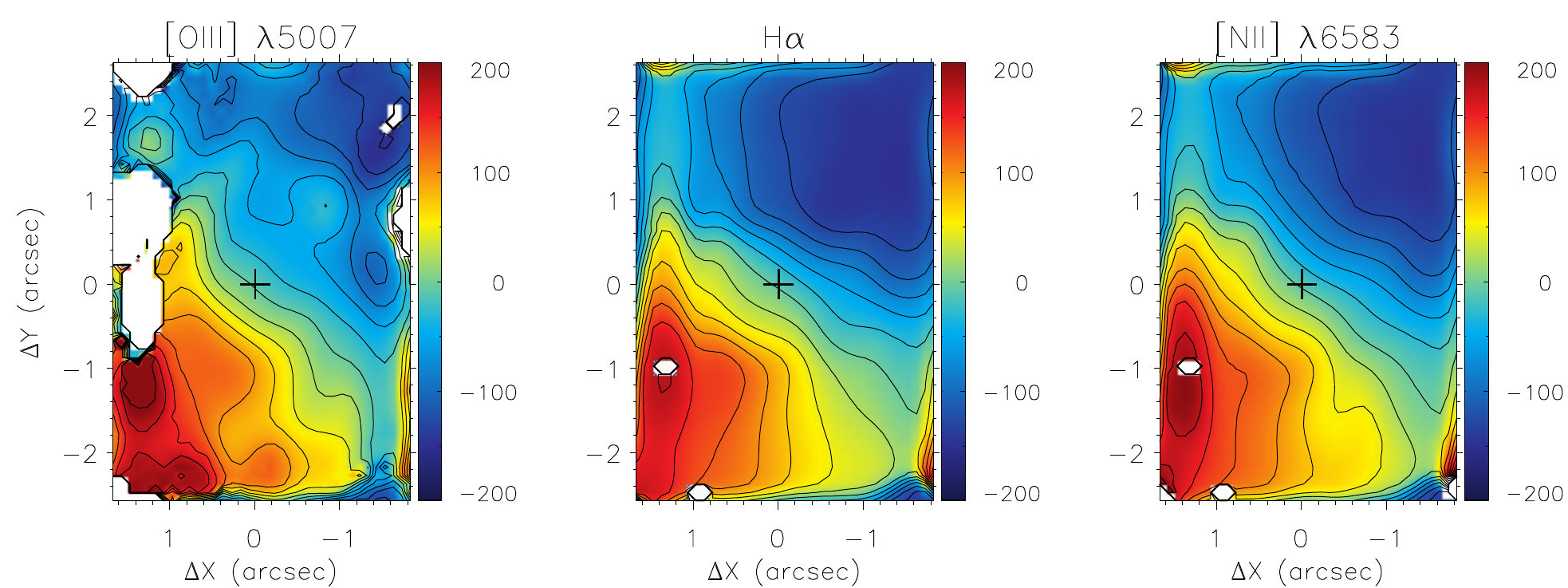}
		\includegraphics[width=\textwidth]{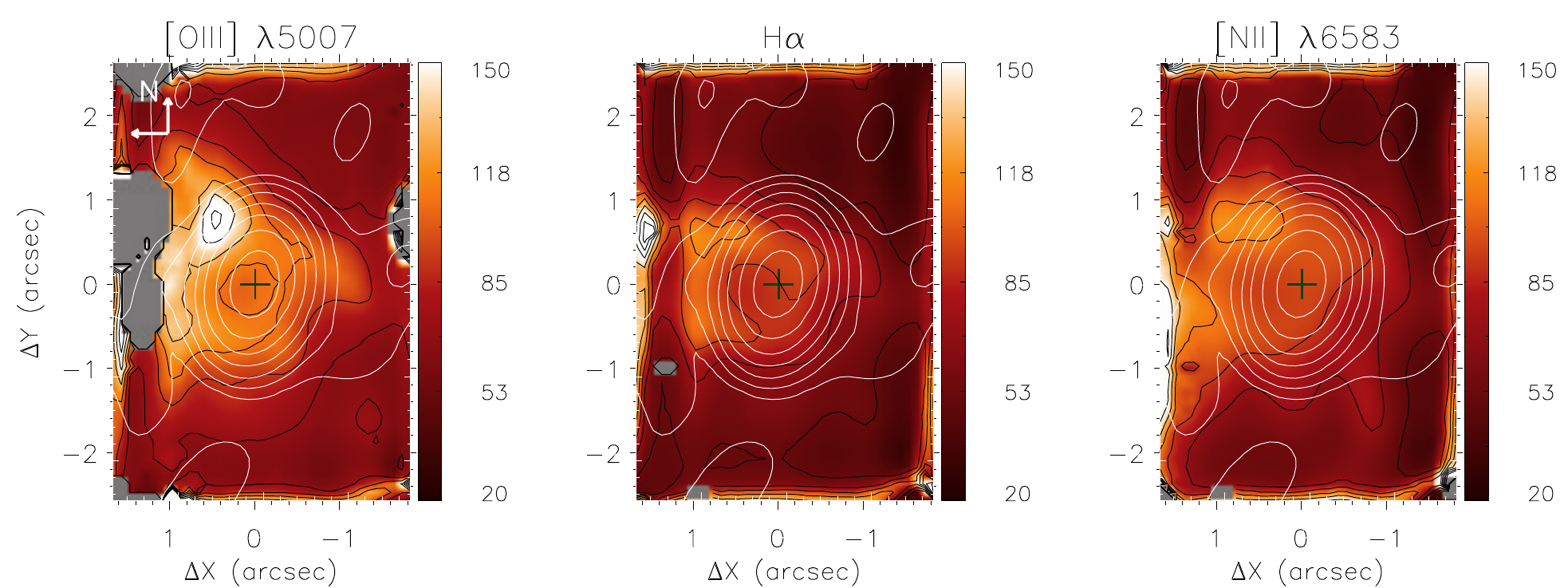}
    \caption{Top panels: line-of-sight velocity fields for the [O\,{\sc iii}] (left), H$\alpha$ (centre) and [N\,{\sc ii}] (right) emitting gas. The color bars show the velocities in units of km\,s$^{-1}$, after the subtraction of the systemic velocity of the galaxy. Bottom panels: velocity dispersion maps for the  [O\,{\sc iii}] (left), H$\alpha$ (centre) and [N\,{\sc ii}] (right) emission-lines, corrected for the instrumental broadening. The white contours presented in the residual map are at 8.5\,GHz. The color bars show the $\sigma$ values in units of km\,s$^{-1}$. The central cross in all panels marks the position of the nucleus}
    \label{fig:velandsig}
\end{figure*}

Figure \ref{fig:velandsig} shows the line-of-sight velocity and velocity dispersion maps for [O\,{\sc iii}], H$\alpha$ and [N\,{\sc ii}] emission lines. As for the stellar velocity field, the velocities are shown relative to the systemic velocity of the galaxy estimated in Sec.~\ref{sec:kinematics}. The velocity fields of all the emission-lines are very similar and present a disturbed rotation pattern with  blueshifts and redshifts observed to the north-west and south-east of the nucleus, respectively, similar to that of the stellar velocity field. The velocity amplitude is nevertheless larger that of the stellar velocity filed, reaching about 200\,km\,s$^{-1}$.

The velocity fields are distorted when compared to a rotation velocity field, and, in the case of H$\alpha$ and [N\,{\sc ii}], the blueshifted side shows a steeper gradient than the redshifted side, while for the [O\,{\sc iii}] the two gradients are more similar, but the velocity field is noisier.

The bottom panels of Figure \ref{fig:velandsig} show the velocity dispersion maps, with values ranging from $\approx$ 40 to 150\,km\,s$^{-1}$ for all emission lines. The highest values are observed for the [O\,{\sc iii}]$\lambda$5007 in the ``triangular-shaped" or ring-shaped region where the highest emission-line ratios are also observed, as discussed in Section\,\ref{sec:lineratios}. This region is also co-spatial with the location of the strongest radio emission, as evidenced by the radio contours from Fig.\,\ref{fig:radio} overplotted on the velocity dispersion maps.

\section{Discussion}

\subsection{Gas excitation}

\subsubsection{Star-forming regions properties}

\begin{table*}
	\caption{Physical properties of the star-forming regions in IRAS11506.}
	\label{tabela}
	\begin{tabular}{ccccc} 
		\hline
		Region & L$_{H_{\alpha}}$ (10$^{39}$ erg s$^{-1}$) &  M (10$^{4}$M$_{\odot}$) & log Q[H$^{+}$] (s$^{-1}$) & SFR (10$^{-3}$M$_{\odot}$ yr$^{-1}$)\\
		\hline
		N & 6 & 1.5 & 50.79 & 5\\
		A & 2 & 0.5 & 50.31 & 2\\
		B & 2 & 0.6 & 50.41 & 2\\
		C & 6 & 0.9 & 50.79 & 5\\
		D & 6 & 0.9 & 50.79 & 5\\
		E & 10 & 2.5 & 51.01 & 8\\
		F & 9 & 2.3 & 50.96 & 7\\
		\hline
	\end{tabular}
\end{table*}

Fig.\,\ref{fig:flux} shows knots of strong emission in H$\alpha$, H$\beta$, [N\,{\sc ii}] and [S\,{\sc ii}], also seen (although with much better angular resolution) in the H$\alpha$+[N\,{\sc ii}] HST images of Fig.\,\ref{fig:hst}, forming a partial ring with on-going star formation.  We extract spectra at the location of the emission-line knots along the ring, labelled as C, D, E and F in Fig.\,\ref{fig:flux} plus one from the nucleus (N) and another two (A and B) between the nucleus and the ring. These spectra were obtained by integrating the fluxes within circular apertures of 0$\farcs$4 radius. We derive the following properties: the H$\alpha$ luminosity, the ionized gas mass, the number of ionizing photons and the star formation rates at the corresponding locations. These properties are listed in table \ref{tabela}.

We first measure the H$\alpha$ fluxes to derive the luminosity of the knots. To obtain the mass of ionized gas we used the following expression from \citet{Peterson1997}:

\begin{equation}\label{eq:mass}
    \frac{M}{\rm M_{\odot}} \approx 2.3 \times 10^{5} \frac{L_{41}(H_{\alpha})}{n_{3}}, 
\end{equation}
where L$_{41}$ is the H$\alpha$ luminosity in units of 10$^{41}$\,erg\,s$^{-1}$ and $n_3$ is the electron density in units of 10$^3$\,cm$^{-3}$. We assume $N_e$=300 cm$^{-3}$, which is the mean value of the electron density of the CNSFRs estimated  from the [S\,{\sc ii}]$\lambda$ 6717/$\lambda$6731 intensity ratio (see also Fig.\,\ref{fig:ratio}). We find values for the mass of ionized gas for each CNSFR ranging from 5$\times$10$^{3}$ M$_{\odot}$ to 2.5$\times$10$^{4}$ M$_{\odot}$. 

In order to obtain the rate of ionizing photons Q[H$^{+}$] for each star-forming region we used the following relation by \citet{osterbrock} and the procedure described in \citet{Heka2018a}:
\begin{equation}
    \left(\frac{Q[H^{+}]}{s^{-1}}\right)=1.03\times10^{12}\left(\frac{L_{H_{\alpha}}}{s^{-1}}\right).
\end{equation}

The Star Formation Rate ($SFR$) was derived under the assumption of continuous star formation regime, using the relation \citep{kennicutt98}:
\begin{equation}
\frac{SFR}{\rm M_\odot\,yr^{-1}}=7.9\times10^{-42}\,\frac{L_{\rm H_\alpha}}{\rm  erg\, s^{-1}}.
\end{equation}

We obtain values of the ionizing photons rate (log Q[H$^{+}$]) ranging from 50.31 to 51.01 and values of SFRs in the range 2 to 8 ($10^{-3}$M$_{\odot}$\,yr$^{-1}$).
The ionizing photons rate values are in agreement with other measurements obtained for nearby Seyfert galaxies \citep[e.g.][]{Wold06,Galliano08,riffel09,riffel2016,hennig2018,Heka2018a,Heka2018b}. 

\subsubsection{Diagnostic diagrams}
\label{sec:diagnostic}

\begin{figure*}
	\includegraphics[width=\columnwidth]{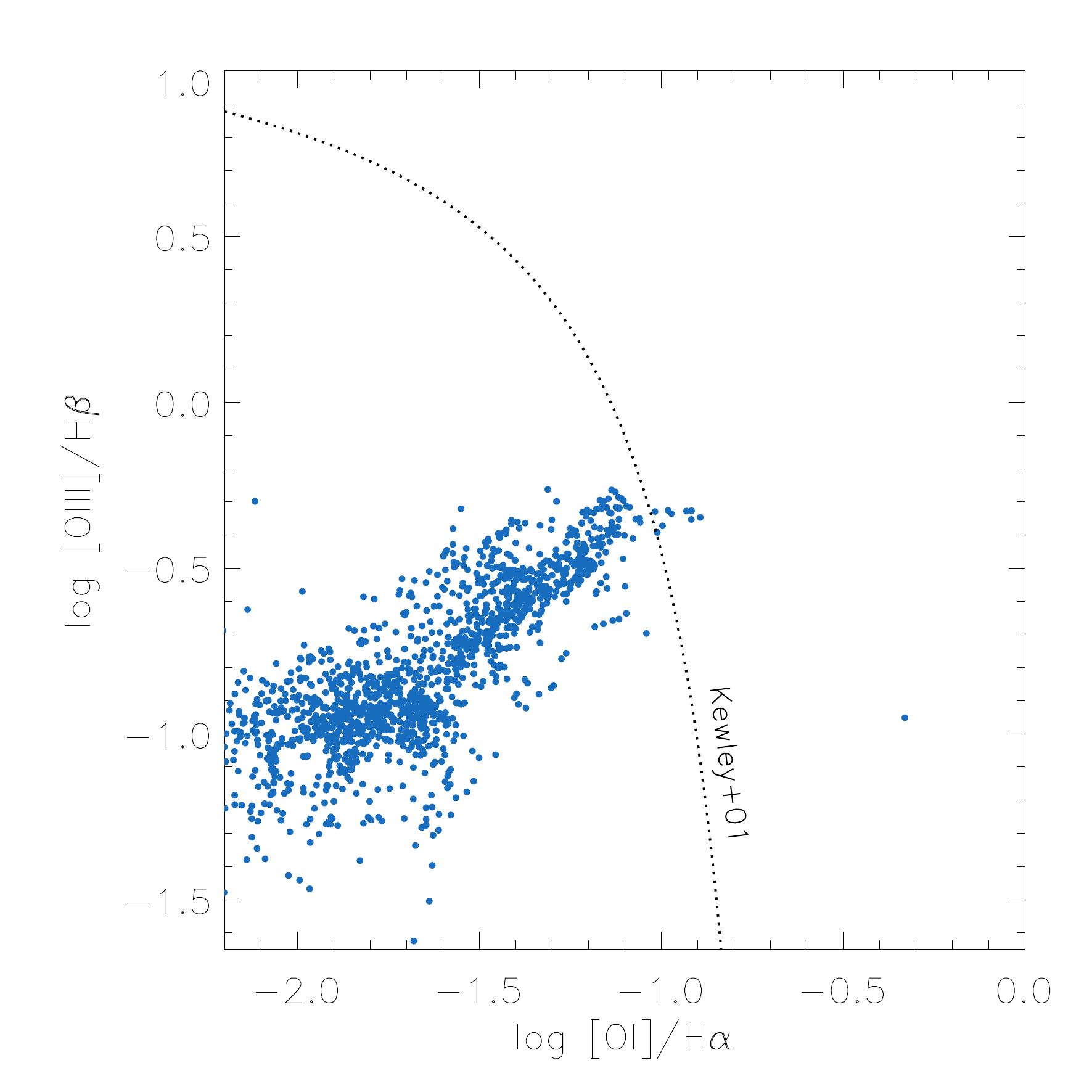}
		\includegraphics[width=\columnwidth]{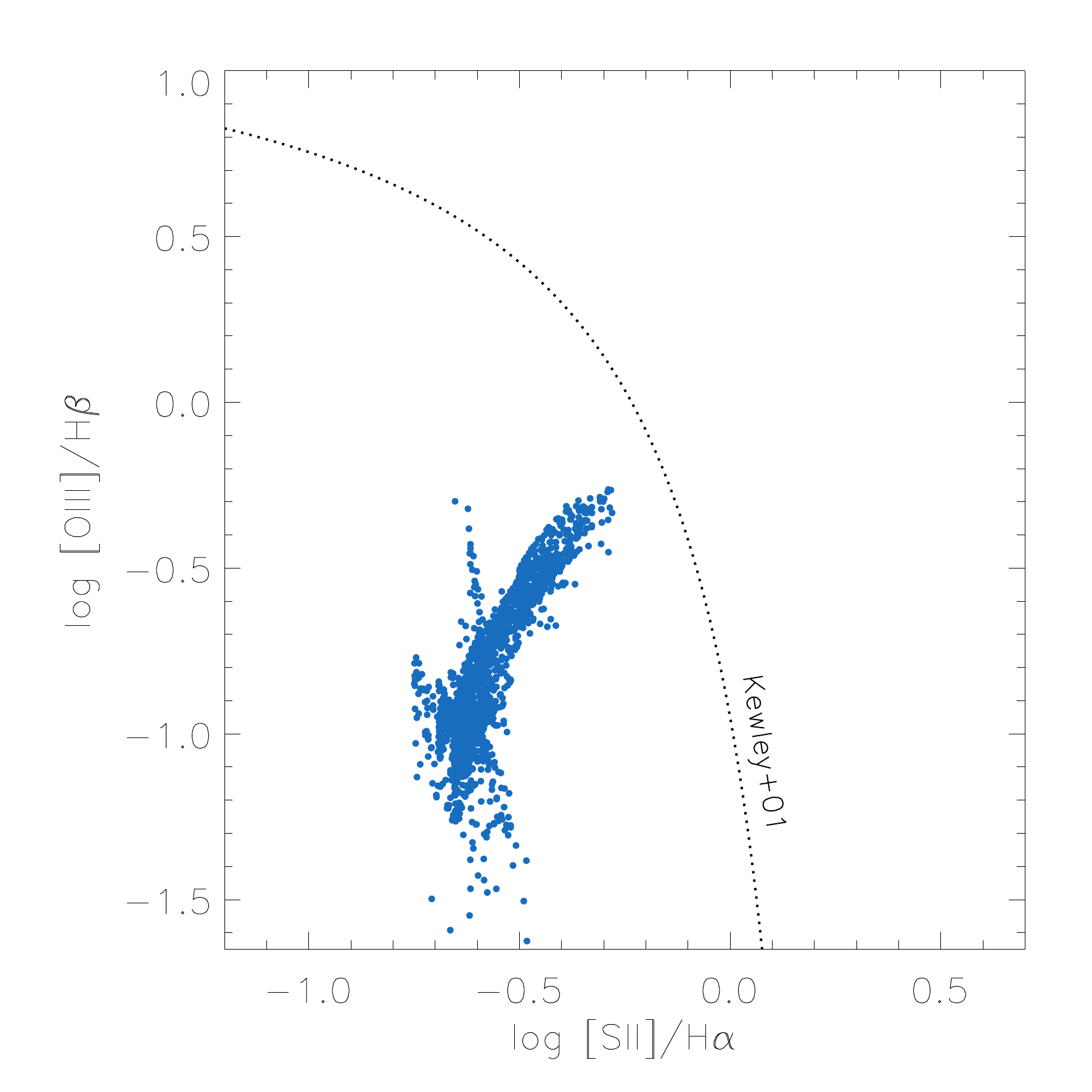}
		\includegraphics[width=\columnwidth]{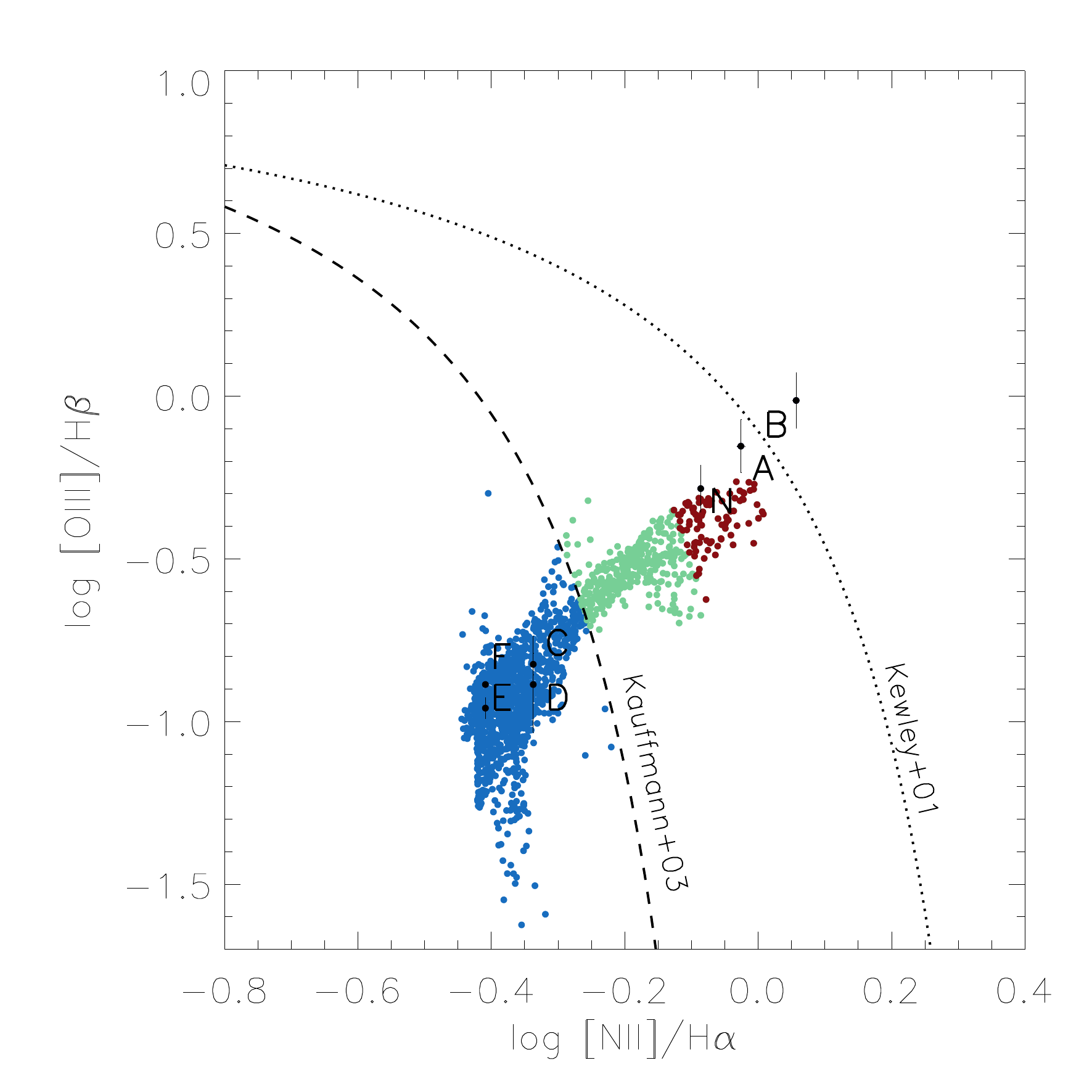}
		\includegraphics[width=\columnwidth]{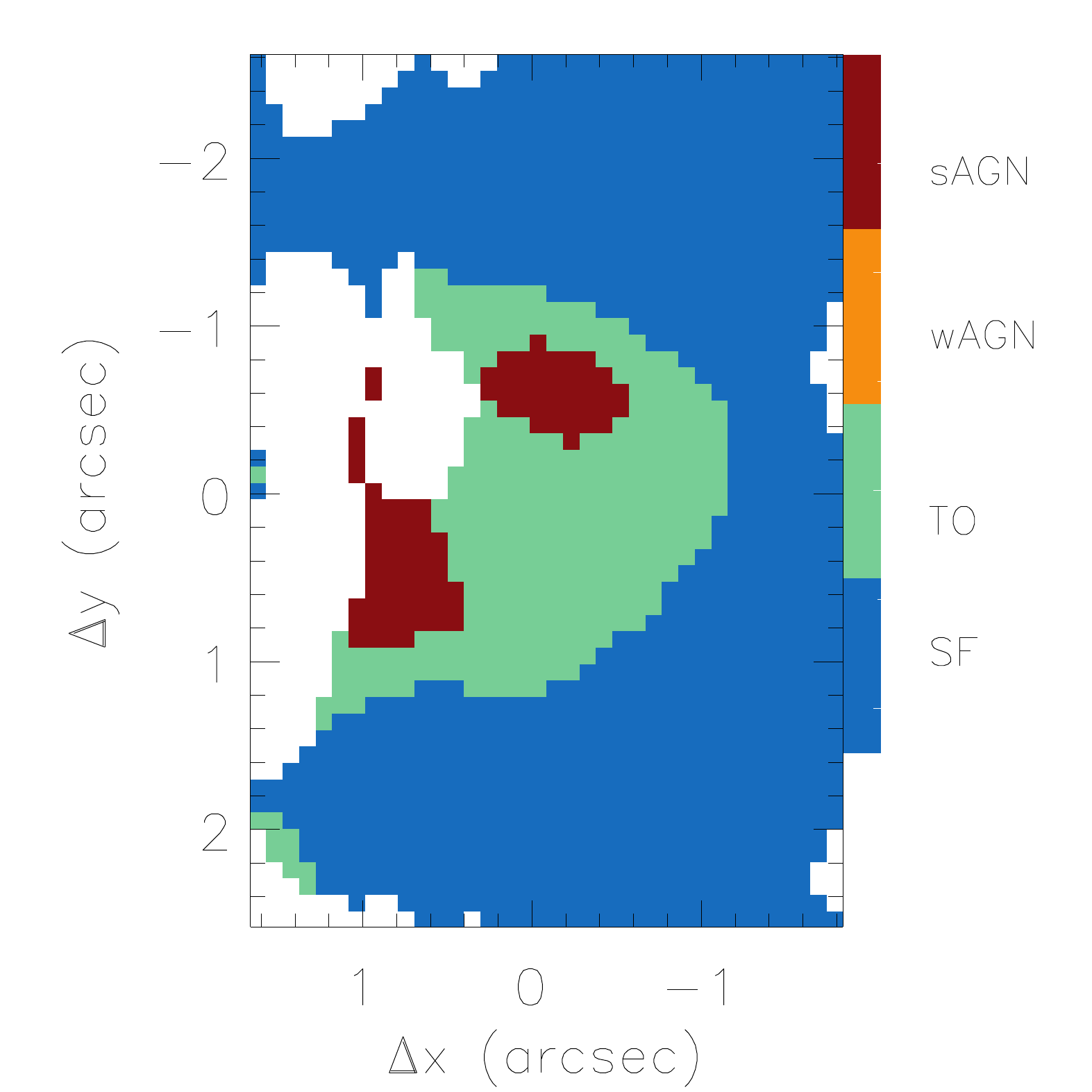}
    \caption{Top panels: [O\,{\sc iii}]$\lambda5007$/H$\beta$ versus [O\,{\sc i}]$\lambda5007$/H${\alpha}$ diagnostic diagram of IRAS11506 (left). [O\,{\sc iii}]$\lambda5007$/H${\beta}$ versus [S\,{\sc ii}]/H${\alpha}$ diagnostic diagram (right).
    Bottom panels: [O\,{\sc iii}]$\lambda5007$/H${\beta}$ versus [N\,{\sc ii}]/H${\alpha}$ diagnostic diagram (left). The dotted and dashed lines represent the Kewley and Kauffmann criteria respectively. Excitation map identifying the regions within the FoV presents distinct excitation mechanisms: strong AGN (sAGN), weak AGN (wAGN), Transition object (TO) and star-forming (SF) (right).}
    \label{fig:bpts}
\end{figure*}

We use the emission-line ratios from the GMOS-IFU data to investigate the nature of the gas emission via the BPT diagnostic diagrams proposed by \citet{baldwin81}. These diagrams allow us to distinguish the gas excitation over the GMOS FoV as characteristic of Starburst, AGN  or composite (hereafter used to identify a region with composite Starburst and AGN excitation).

We present three diagnostic diagrams and an excitation map presented in Figure \ref{fig:bpts}, where each spaxel within the FoV corresponds to a point in these diagrams. The dashed curve is from \citet{Kauffmann2003} and the dotted one from \citet{Kewley2001} which have been used to distinguish ionization due to Starburst, AGN and composite.

The top panels of Fig.~\ref{fig:bpts} show the [O\,{\sc iii}]/H$\beta$ versus [O\,{\sc i}]/H${\alpha}$ and the [O\,{\sc iii}]/H${\beta}$ versus [S\,{\sc ii}]/H${\alpha}$ diagrams. Almost all spaxels are to the left of the Kewley's line, indicating that the gas excitation is dominated by Starburst activity and just a few spaxels show excitation that could be due to an AGN. The bottom panels of Fig.\,\ref{fig:bpts} show the [O\,{\sc iii}]/H${\beta}$ versus [N\,{\sc ii}]/H${\alpha}$ diagnostic diagram in the left and the corresponding excitation map in the right panel. The [O\,{\sc iii}]/H${\beta}$ versus [N\,{\sc ii}]/H${\alpha}$ diagnostic diagram shows points both in the region to the left of the Kauffmann's curve indicating Starburst excitation and in the region between Kauffmann's and Kewley's lines, indicating a composite origin for the observed gas emission.
We split the composite region in ``sub-regions", one with higher excitation (identified by the dark red points) and another with lower excitation (in light green).

The capital letters N, A, B, C, D, E and F in the [O\,{\sc iii}]/H${\beta}$ versus [N\,{\sc ii}]/H${\alpha}$ diagram show the location of the regions identified in Fig. \ref{fig:flux}. While the regions N, A and B are observed in the region that represents composite ionization/excitation
the C, D, E and F regions from the ring are observed in the Starburst region, in agreement with the interpretation that they are star-forming regions.

The bottom right panel of Fig.\,\ref{fig:bpts} shows that the excitation within the bubble of $\approx$240\,pc radius is higher than the starburst excitation of the 500\,pc ring, favoring the presence of mixed AGN and starburst excitation in the former region. But we note that the highest excitation is observed not at the nucleus but in regions at $\approx$\,200\,pc from it to the SE, E and NW of the nucleus, at the ``borders" of the central bubble. This suggests that shocks are probably responsible to this higher excitation, as supported also by the enhanced gas velocity dispersion in these regions, and discussed further in the next section.

We have checked if shock model sequences, such as those of \citet{Allen+08} that have been successfully used to reproduce similarly enhanced [N{\sc ii}]/H$\alpha$ and [S{\sc ii}]/H$\alpha$ line-ratio values in another OHM galaxy of our sample -- IRAS\,17526+3253, in \citet{Dinalva2019}, could reproduce the above line ratios in IRAS\,11506. We concluded that, although the sequence of these line ratios can indeed be reproduced by these shock models, for velocities in the range 100--300\,km\,s$^{-1}$, the [O{\sc iii}]/H$\beta$ ratio values of IRAS\,11506 are too low to be reproduced by such models. Our conclusion is that -- although it is clear that shocks are present, they are not the main excitation mechanism of the gas, that seems to be dominated by Starburst excitation, according to the BPT diagrams.

\subsection{Gas and stellar kinematics}
\label{sec:kinematics}


\begin{figure*}
	\includegraphics[width=\textwidth]{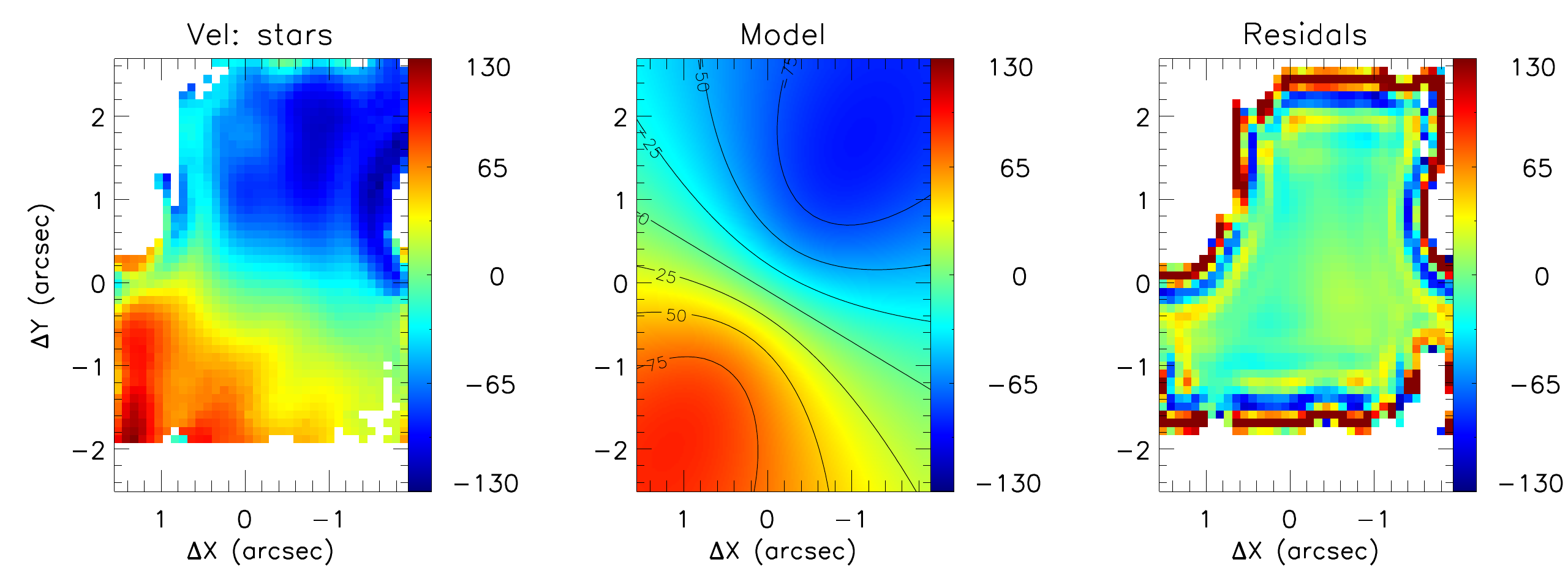}
    \caption{Observed H$\alpha$ stellar velocity field (left), rotating disc model (centre) and residual map (right), obtained as the difference between the observed velocities and the model.}
    \label{fig:modelo}
\end{figure*}

\begin{figure*}
	\includegraphics[width=\textwidth]{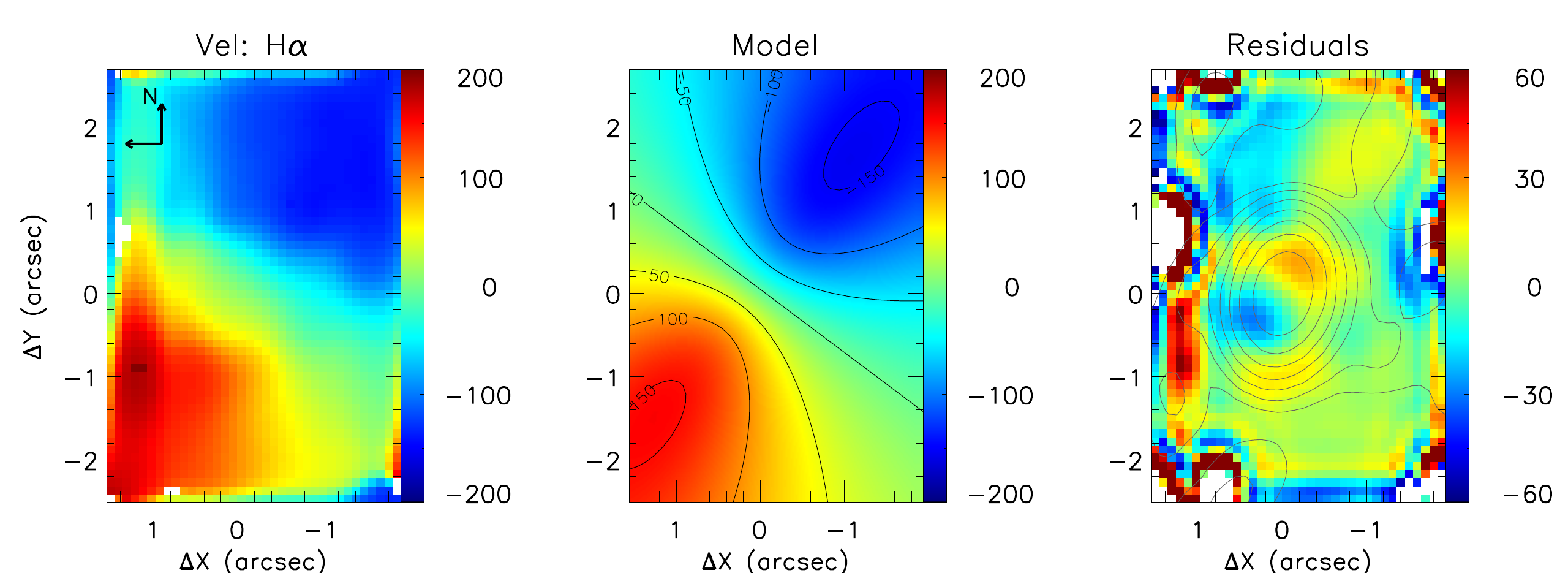}
    \caption{Observed H$\alpha$ velocity field (left), rotating disc model (centre) and residual map (right), obtained as the difference between the observed velocities and the model. The contours presented in the residual map are at 8.5\,GHz.
	The central cross marks the position of the nucleus}
    \label{fig:modeloha}
\end{figure*}

In an effort to understand the stellar and gas kinematics of IRAS\,11506 we fit the corresponding velocity fields with a rotating disc model. We use an analytical model under the assumption that the stars and gas have circular orbits in the plane of the galaxy \citep{kruit78,bertola91}, and that the circular velocity is given by

\[ 
 V_{mod}(R,\psi)=v_{s}+ 
\]
\begin{equation} \label{eq-bertola}
    \frac{AR\cos(\psi-\psi_{0})\sin(i){\cos^{p}(i)}}{\{R^{2}[\sin^{2}(\psi-\psi_{0})+\cos^{2}(i)\cos^{2}(\psi-\psi_{0})]+{c_{0}}^{2}\cos^{2}(i)\}^{\frac{p}{2}}}, 
    \end{equation}
    
\noindent where $R$ is the radial distance from the nucleus projected on the plane of the sky and $\psi$ is its corresponding position angle. The systemic velocity of the galaxy is given by $v_{s}$, $A$ is the velocity amplitude, $\psi_{0}$ is the position angle of the line of nodes, $c_{0}$ is a concentration parameter, defined as the radius where the rotation curve reaches 70\% of the velocity amplitude and $i$ is the disc inclination relative to the plane of the sky. The $p$ parameter measures the slope of the rotation curve beyond the maximum amplitude, where, usually 1\,$\leq$\,$p$\,$\leq$\,$\frac{3}{2}$. For $p$=1, the rotation curve at large radius is asymptotically flat while for $p$=$\frac{3}{2}$ the system has a finite mass. 

We fit the stellar and gas velocity fields by the equation above using the {\sc mpfitfun idl} routine \citep{mark09} that performs a non-linear least-squares fit. During the fit we fixed the kinematic centre to the position of the continuum peak and the $p$ parameter to $p$=1.5, as we are probing the inner region of the galaxy. The H$\alpha$ velocity field was chosen to represent the gas kinematics because H$\alpha$ is the brightest emission-line over most of the FoV, being also similar to the velocity fields of the other emission lines.

Figure \ref{fig:modelo} shows the observed stellar velocity field (left), best fit rotation disc model (centre) and the residual map (right) obtained as the difference between the first and the latter. The velocity residuals are small at all locations (except at the borders of the regions where the uncertainties in the measured velocities are high), indicating that the stellar velocity field is well reproduced by a rotating disc. The resulting parameters for the best fitted model are: $v_s=3076\pm5$ km\,s$^{-1}$, $A =283\pm14$~km\,s$^{-1}$, $c_0=$1\farcs6$\pm$0\farcs1, $\psi_0=148^\circ\pm3^\circ$ and $i=41^\circ\pm5^\circ$.
\citet{Santaella2016} used a similar kinematic model to fit the velocity field obtained from the first moment of the CO(2-1) emission. They obtained $v_{s}=3080\pm4$ km\,s$^{-1}$, $\psi_{0}$=133$^\circ\pm2^{\circ}$ and $i=43^\circ$ and so our results are in good agreement with theirs.

Figure \ref{fig:modeloha} presents the observed H$\alpha$ velocity field (left), the rotating disc model (centre) and residual map (right) between these two. 

Neglecting the regions of high residuals at the borders of the FoV, which are more affected by noise, the residual map shows some structures that suggest the presence of non-circular motions. In particular, within $\approx$1$^{\prime\prime}$ from the nucleus, there are residual redshifts to the north-west, with velocities of $\approx$\,30\,km\,s$^{-1}$ and residual blueshifts to the opposite direction, south-east, with similar velocities. The resulting parameters for the best fitted model are: $v_s=3075\pm1$ km\,s$^{-1}$, $A =365\pm6$~km\,s$^{-1}$, $c_0=$1\farcs4$\pm$0\farcs4, $\psi_0=125^\circ\pm3^\circ$ and $i=54^\circ\pm5^\circ$.

We note that the excess blueshifts and redshifts close to the nucleus along PA=-45/130$^{\circ}$ are oriented in the direction of spots of increased line ratios of [O\,{\sc iii}]/H$\beta$ and [S\,{\sc ii}]/H$\alpha$ in Fig.\,\ref{fig:ratio}, coinciding also with the orientation of the elongation seen in the outer contours of the spectral index radio image of Fig.\,\ref{fig:spectralindex}. In addition, the radio spectral index image shows approximately constant values in this elongated structure. All these features suggest the presence of a weak bipolar outflow at PA=-45/130$^{\circ}$. 

The gas velocity dispersion maps of Fig.\,\ref{fig:velandsig} also show increased values along the above PA, but are higher than the surroundings within the whole inner bubble, with the highest values also observed towards the borders of the 240\,pc bubble and in particular towards its eastern border, where the [O\,{\sc iii}]/H$\beta$ ratio is also highest. The whole region of increased velocity dispersion coincides with the region of strongest radio emission.

We conclude that, although the velocity residuals suggest a modest outflow along PA=-45/130$^{\circ}$, the most robust signature of feedback from the AGN are the enhanced velocity dispersions towards the borders of the inner bubble suggesting interaction between the plasma bubble and the surrounding medium, pushing and shocking it and producing higher line ratio values.

\subsection{The nature of the nuclear emission: is IRAS11506 hiding an AGN?}

As discussed in Sec.\,\ref{sec:diagnostic}, we find enhanced gas excitation within the inner 240\,pc bubble of IRAS\,11506, suggesting the presence of a weak AGN ionizing/exciting the gas there, mixed with Starburst ionization/excitation. 

The presence of an additional source of excitation besides the young stars of a Starburst is also supported by the presence of radio emission in the region as revealed by the radio contours in Fig.\,\ref{fig:velandsig}, and the enhanced gas velocity dispersion towards the borders of the bubble where the excitation is also highest. The enhanced velocity dispersion supports the presence of shock excitation, as discussed in the previous section, most probably due to the interaction of the radio bubble with the ambient gas. 

The VLA data radio core spectral index is compatible with an AGN \citep[e.g.][]{Sadler95,Roy2000,Orienti2010,Kharb2010,Bontempi2012}. This implies optically-thin synchrotron emission that could nevertheless also originate in a nuclear starburst. The elongation seen in the spectral index map (Fig.\,\ref{fig:spectralindex}), on the other hand, makes the radio core in IRAS11506 to resemble more an AGN radio jet than a nuclear starburst.

The discussion above on the excitation and radio emission suggests that the gas in the inner 240\,pc (the inner bubble) is mostly ionised by starbursts in the region. The enhanced velocity dispersion associated with a plasma bubble withing the inner 240\,pc and a possible bipolar outflow supports also the presence of a weak AGN there. We are probably witnessing the birth of an AGN that just emitted a bubble of plasma that is beginning to interact with the surrounding gas via mostly shocks at the present stage.

\subsection{Comparison with previous studies}

We now compare our results for IRAS\,11506 with those recently obtained by previous authors and also with the results of our previous similar studies of OHM galaxies. We have been performing a multiwavelength investigation of the excitation mechanism of these galaxies with the goal of relating the merger state and OH maser properties to Starburst and AGN activity.

\subsubsection{Recent results from the literature}

\citet{Cazzoli2014} studied the multi-phase component structure and wind properties of IRAS\,11506 based on IFS obtained with the instruments VIMOS and SINFONI of the Very Large Telescopes (VLT). Their data provide information about kinematic and dynamical properties of the stellar component and gas phases using as tracers the NaD absorption doublet, H$\alpha$, and the CO bands in the near-infrared. The stellar kinematics they derive from the CO bands is similar to the gas kinematics they derive from H$\alpha$, showing also a larger rotation amplitude for the gas (203$\pm$4 km\,s$^{-1}$) than for the stars (188$\pm$11 km\,s$^{-1}$). 

However, the resolution of the maps of \citet{Cazzoli2014} based on the optical data is about 1 arcsec, and, although their Fig. 5 shows an increase in the velocity dispersion at the nucleus, this is observed in only 3-4 pixels. The sampling of their optical IFS observations is 0$\farcs$7 per fiber, while the seeing in the observations was reported to be 0$\farcs$9. As judging from these numbers and from their Fig. 5, with their data they could not sample the structure we see in our Fig.\,\ref{fig:velandsig} that resolves the enhanced $\sigma$ bubble we have found in our maps. 
We note also that the regions with highest $\sigma$ in our [O\,{\sc iii}] maps of  Fig.\,\ref{fig:velandsig} are seen at approximately 1 arcsec to the north-east. This is the direction where \citet{Cazzoli2014} report the presence of an outflow seen in neutral gas via the observations of the NaD optical absorption feature. Their Fig. 4 shows that the outflow -- observed reaching blueshifts of up to $\approx$150\,km\,s$^{-1}$ is seen approximately between 3 and 6 arcsec, which lies beyond the borders of our FoV. Although \citet{Cazzoli2014}  argue that the outflow is due to feedback from star formation, our more detailed sampling of the region, suggests that the origin of the outflow may be an AGN. 

In a following work, \citet{Santaella2016} used high spatial resolution ($\sim$ 60 pc) ALMA CO (2-1) observations, HST optical and near-IR images, as well as VLT/SINFONI IFS data to characterize the resolved outflow in IRASF\,11506, finding that the ALMA observations revealed a number of cold molecular gas knots along the direction of the outflow seen in neutral gas by \citet{Cazzoli2014}. They concluded that an extremely obscured nuclear starburst produces the outflow. We favor instead the presence of a faint/recently born AGN at the nucleus due to the resolved structure of the 240\,pc bubble and enhanced line ratios correlated with the radio structure. 

One possible scenario to reconcile the above results with ours is that we are probing the nuclear region which is revealing a new ejection of a plasma bubble that is pushing the surrounding gas, thus increasing its velocity dispersion. Mostly shocks but possibly also some escape of AGN radiation could then explain the increase in the emission-line ratios seen in Fig.\,\ref{fig:ratio}. A new ejection towards the north-east may also be occurring, which could explain the increased velocity dispersion towards this direction \citep{Kharb2017}.

\subsubsection{Comparison with our previous studies of OH Megamaser galaxies}

In the first paper of this series of multiwavelength studies of OHM galaxies, \citet{Dinalva2015} presented a study of IRAS\,16399-0937 which is characterized as a mid to advanced merger with two nuclei embedded in a diffuse envelope. The nuclei are separated by 3.4\,kpc and the spectral energy distribution revealed that the northwestern nucleus hosts a dust embedded AGN of luminosity $L_{bol}$ $\approx$ 10$^{44}$ erg s$^{-1}$.

In the second study, \citet{Heka2018a} used GMOS-IFU, HST and VLA data of IRAS\,23199+0123, which is an interacting pair of galaxies separated by 24 kpc. We were able to detect a Seyfert 1 nucleus in the western member of the pair revealed by the presence of an unresolved broad H$\alpha$ line. From the VLA data we were able to report a new maser detection, determine the position of the masers and conclude that the masers sources are correlated to shocks driven by AGN outflows.

The third study was performed for IRAS\,03056+2034 \citep{Heka2018b}, a barred spiral galaxy with irregular structures that could indicate a past interaction and also reveals a circumnuclear ring of star-forming regions and floculent spiral arms. The BPT diagram and VLA data suggest the presence of an embedded AGN in the midst of by star-forming regions surrounding the nucleus.

The fourth study was on IRAS\,17526+3253 \citep{Dinalva2019}. This galaxy is in mid-stage major merger with the nuclei separated by $\approx$ 8.5 kpc that hosts both OH and H$_{2}$O masers. There is no strong evidence of the presence of an AGN, although this possibility can not be excluded based on the analysis of the presented data. 

And finally, here we presented a study for IRAS\,11506-3851, which is an isolated spiral galaxy that does not show clear signs of interaction, but shows a double-barred structure. As in the other galaxies, star-forming regions surround the nucleus. Moreover, the analysis of the VLA data, BPT diagrams and gas kinematics suggest the presence of a weak AGN immersed in a region dominated by star formation. 

In summary, in the 5 OHM galaxies studied so far, the most common pattern is the presence of plenty of star-forming regions. Three are clear mergers, and in two of them we clearly find an AGN. The other two seem to be isolated galaxies, but both are barred, with one of them showing asymmetries that could be related to a past merger. Both show signatures of an embedded AGN in the midst of star-forming regions surrounding the nucleus. 

We note that, although the conventional thinking is that the megamaser relates to starburst activity, our on-going observations of OHM galaxies show that the megamaser can also be related to the presence of AGN activity. We aim to investigate a possible relation between the luminosity of the megamaser and that of the AGN, but our sample of 5 galaxies is still small for this. The AGN-related luminosities we have so far for all the targets are the megamaser and the H$\alpha$ luminosities, with the caveat that it is not easy to separate the H$\alpha$ luminosity of the AGN from that of the surrounding star formation, that is probably 
``contaminating" the central H$\alpha$ luminosity. The only significant trend is that the weakest Maser observed so far is in IRAS\,11506, which also has the lowest nuclear H$\alpha$ luminosity of the sample we have observed.


\section{Conclusions}

We have performed a multiwavelength data analysis of the OHM galaxy IRAS\,11506-3851 comprising HST and VLA images and GMOS-IFU spectroscopy. The GMOS-IFU observations cover the inner 6.7 $\times$ 9.6 kpc$^{2}$ at a spatial resolution of 193\,pc and spectral resolution of 1.8\AA. Our main conclusions are:
\begin{itemize}
    \item The HST images reveal an isolated spiral galaxy that combined with the GMOS-IFU observations allowed us to identify a partial circumnuclear ring of star formation with $\approx$\,500\,pc radius and an obscured region to the north-east of the nucleus.
    
    \item From the diagnostic diagrams we show increased gas excitation internal to the star-forming ring, within a radius of 240\,pc from the nucleus, due to a mixed contribution of Starburst and AGN excitation, apparently mostly due to shocks, while the excitation in the rest of the FoV, dominated by the 500\,pc ring, is due to Starbursts. 
    
     \item The VLA images reveal a steep spectrum kpc-scale radio source co-spatial with the 240\,pc radius region of enhanced excitation that we attribute to a plasma bubble emitted by a faint AGN.
    
    \item The VLA data radio core spectral index is compatible with an AGN and the elongation observed in the spectral index map makes the radio core in this galaxy to resemble more an AGN than a nuclear Starburst. 
    
    \item The residuals of the fit of a rotating-disc model to the gas kinematics reveal mild non-circular motions suggesting an outflow within the inner 240\,pc co-spatial to one previously seen in neutral gas. Moreover, the whole region presents higher velocity dispersion than the surroundings, that are co-spatial with the radio emission. The enhanced velocity dispersion at the borders of this region, co-spatial with regions of increased excitation suggest shock excitation by the plasma bubble presumably emitted by a nuclear AGN.
    
    \item From the H$\alpha$ fluxes we estimated physical properties of the star forming regions of the 500\,pc ring. The masses of ionized gas are in the range of (0.1 - 0.8) $\times$ 10$^{4}$ M$_{\odot}$. The ionizing photons rate log(Q[H$^{+}$]) is in the range 50.31 - 51.01 (units of photons\,s$^{-1}$) and the average star formation rate is about 0.005 M$_{\odot}$yr$^{-1}$.

\end{itemize}

We argue that our observations of this OHM galaxy is unveiling the ejection of a plasma bubble that is pushing the surrounding gas and either shocks and ionizing photons from a faint AGN can explain the increased excitation of the region. We thus suggest that the origin of the outflows previously reported in neutral and molecular gas could be this AGN. 

This is the fifth OHM galaxy from a sample of 15 that we analyse using similar data, whose unresolved nuclear gas emission was known to be dominated by star formation. And is the fourth one for which GMOS-IFU data has resolved the presence of a previously unknown AGN at the nucleus. These results support the hypothesis that OHM galaxies harbor faint AGN being triggered by recent accretion of matter to their central SMBH.

\section*{Acknowledgements}

This work is based on observations obtained at the Gemini Observatory, which is operated by the Association of Universities for Research in Astronomy, Inc., under a cooperative agreement with the NSF on behalf of the Gemini partnership: the National Science Foundation (United States), National Research Council (Canada), CONICYT (Chile), Ministerio de Ciencia, Tecnolog\'{i}a e Innovaci\'{o}n Productiva (Argentina), Minist\'{e}rio da Ci\^{e}ncia, Tecnologia e Inova\c{c}\~{a}o (Brazil), and Korea Astronomy and Space Science Institute (Republic of Korea). 
This research has made use of NASA's Astrophysics Data System Bibliographic Services. This research has made use of the NASA/IPAC Extragalactic Database (NED), which is operated by the Jet Propulsion Laboratory, California Institute of Technology, under contract with the National Aeronautics and Space Administration. Support for programme HST-SNAP 11604 was provided by NASA through a grant from the Space Telescope Science Institute, which is operated by the Association of Universities for Research in Astronomy, Inc., under NASA contract NAS 5- 26555. RAR thanks support from Conselho Nacional de Desenvolvimento Cient\'ifico e Tecnol\'ogico (202582/2018-3, 304927/2017-1 and 400352/2016-8) and Funda\c c\~ao de Amparo \`a pesquisa do Estado do Rio Grande do Sul (17/2551-0001144-9 and 16/2551-0000251-7). CH thanks CAPES for financial support.

\section*{DATA AVAILABILITY}

The GEMINI data used in this work is publicly available online via the GEMINI archive https://archive.gemini.edu/searchform/, with project code GS-2014A-Q-75. 
The VLA data is available at https://science.nrao.edu/facilities/vla/archive and project codes are AB660 and AL508.
Finally, the HST data is available at https://archive.stsci.edu/hst/ with project code 11604.
The maps produced from these data can be shared on reasonable request to the corresponding author.













\bsp	
\label{lastpage}
\end{document}